\documentclass[final,5p,times,twocolumn]{elsarticle}

\usepackage{lineno}
\usepackage{graphicx}
\usepackage{amssymb}
\usepackage{pdflscape}
\usepackage[english]{babel}
\usepackage[dvipsnames,svgnames,x11names]{xcolor}
\usepackage{booktabs,tabularx}
\usepackage{multirow}
\usepackage{subfig} 
\usepackage{hyperref}
\usepackage{amsmath}
\usepackage{commath}
\usepackage[english]{babel}
\usepackage[ruled]{algorithm2e}
\SetEndCharOfAlgoLine{}
\usepackage[leftmargin=6em,rightmargin=12em,indentfirst=false]{quoting}
\usepackage{siunitx}

\usepackage[final]{changes}
\usepackage{float}

\usepackage{lineno}
\usepackage{tikz}
\usepackage{comment} 
\usepackage[export]{adjustbox}

\usepackage{graphicx}
\usepackage[utf8]{inputenc}
\usepackage[export]{adjustbox}
\usepackage{wrapfig}
\usepackage{dcolumn}

\makeatletter
\newcolumntype{T}[3]{>{\textfont0=\the@{#1}{#2}{#3}}c<{\DC@end}}
\makeatother

\usepackage{pgfplots}
\pgfplotsset{width=10cm,compat=1.9}

\usepackage{array}

\newcolumntype{L}[1]{>{\raggedright\let\newline\\\arraybackslash\hspace{0pt}}m{#1}}
\newcolumntype{C}[1]{>{\centering\let\newline\\\arraybackslash\hspace{0pt}}m{#1}}
\newcolumntype{R}[1]{>{\raggedleft\let\newline\\\arraybackslash\hspace{0pt}}m{#1}}

\usepackage{subfiles}

\usepackage{todonotes}

\journal{Applied Energy}
\begin{document}
	
\begin{frontmatter}

\title{EnergyStar++: Towards more accurate and explanatory building energy benchmarking}

\author{Pandarasamy Arjunan$^a$, Kameshwar Poolla$^{a,b}$, Clayton Miller$^{c,*}$, }		

\address{$^a$Berkeley Education Alliance for Research in Singapore (BEARS), Singapore}
\address{$^b$University of California, Berkeley, United States}
\address{$^c$Building and Urban Data Science (BUDS) Lab, Dept. of Building, School of Design and Environment (SDE), National University of Singapore (NUS), Singapore}
\address{$^*$Corresponding Author: clayton@nus.edus.sg, +65 81602452}

\begin{abstract}
Building energy performance benchmarking has been adopted widely in the USA and Canada through the \emph{Energy Star} Portfolio Manager platform. Building operations and energy management professionals have long used this simple \emph{1-100} score to understand how their building compares to its peers. This single number is easy to use but is created by potentially inaccurate multiple linear regression (MLR) models and lacks much further information about \emph{why a building achieves that score}. This paper proposes a methodology that enhances the existing Energy Star calculation method by increasing accuracy and providing additional model output processing to help \emph{explain why} a building is achieving a particular score. Two new prediction models were proposed and tested: multiple linear regression with feature interactions (MLRi) and gradient boosted trees (GBT). Both models performed better than a baseline Energy Star MLR model as well as four baseline models from previous benchmarking studies. This paper shows that for six building types, on average, the third-order MLRi models achieved a 4.9\% increase in adjusted $R^2$ and a 7.0\% decrease in normalized root mean squared error (NRMSE) over the baseline MLR model. More substantially, the most accurate GBT models, on average, achieved a 24.9\% increase in adjusted $R^2$ and a 13.7\% decrease in NMRSE against the baseline MLR model. In addition, a set of techniques was developed to help determine which factors most influence a building's energy use versus its peers using SHapley Additive exPlanation (SHAP) values. The SHAP force visualization, in particular, offered an accessible overview of the aspects of the building that influenced the score that even non-technical users can interpret. This methodology was tested on the 2012 Commercial Building Energy Consumption Survey (CBECS)(1,812 buildings) and public data sets from the energy disclosure programs of New York City (11,131 buildings) and Seattle (2,073 buildings).
\end{abstract}

\end{frontmatter}



\section{Introduction}
\label{sec:introduction}    
The benchmarking of a non-residential building is the process of measuring its energy performance in relation to its peers in order to identify inefficient behavior. Building performance benchmarking (also known as rating or labeling) systems are becoming central in the evaluation of the energy performance of buildings. Up to 40\% of the United States commercial building stock is benchmarked on the Energy Star Portfolio Manager platform including buildings from over half of Fortune 100 companies, half of the largest U.S. healthcare organizations, and sometimes even buildings from entire cities\footnote{\url{https://www.energystar.gov/buildings/facility-owners-and-managers/existing-buildings/use-portfolio-manager}}. In the European Union (EU), the 2010 Energy Performance of Buildings Directive\footnote{\url{https://ec.europa.eu/energy/en/topics/energy-13efficiency/energy-performance-of-buildings}} mandated that all member nations implement building energy labeling schemes that provide ratings to buyers in the real estate market to evaluate energy performance \cite{annunziata_towards_2013}. Other parts of the world have used these systems as inspirations for their own building performance benchmarking activities including China \cite{bank_china_2013}, Singapore \cite{singapore_building_and_construction_authority_bca_bca_2018, dong_holistic_2005, lee_building_2008}, and Australia \cite{bannister_nabers:_2012}. These systems have gained traction over the years as a means of evaluating the general performance of a building as compared to its peers; this process is especially useful in planning retrofits or other energy savings interventions. Benchmarking as a concept is referred to by different terminologies including \emph{energy labelling}, \emph{energy certification}, \emph{energy rating}, \emph{asset rating}, \emph{operational rating}, and \emph{O\&M rating} \cite{perez-lombard_review_2009, goldstein_classification_2014}. 

The widespread deployment of these types of rating systems has had a significant impact on the energy efficiency of the building stock. For example, in the USA, a 2012 study by the Environmental Protection Agency (EPA) examined 35,000 buildings that had undergone Energy Star benchmarking and found over a 7\% decrease in energy use over a four year period \cite{us_environmental_protection_agency_epa_u.s._2012}.

\subsection{Growth of city-scale energy disclosure programs}
In addition to building energy benchmarking systems at the national level, numerous cities and states have started to require non-residential buildings to disclose their monthly energy consumption. The first energy benchmarking ordinance was passed in the city of Washington, D.C., in 2008. Since then, 27 other cities and three states in the United States have implemented similar policies \cite{institute_for_market_transformation_comparison_2019}. These ordinances mandate that a portion of a city’s building stock must undergo energy benchmarking and must disclose such data to the public. The growing popularity of such policies is unsurprising, given its value and importance. 

In New York City, the local energy benchmarking program was estimated to reduce energy use by 14\% over four years, with cumulative energy savings exceeding \$267 million \cite{city_of_new_york_new_2017}. Another more in-depth analysis showed that the New York City energy disclosure program resulted in mandatory energy audits that had a 2.5\% and 4.9\% energy reduction impact for residential and office buildings, respectively \cite{kontokosta2020impact}. A comprehensive study of 24 state and city jurisdictions stated that ``all but one of the B\&T (benchmarking and transparency) policy evaluation studies reviewed for this report indicate some reduction (from 1.6 to 14 percent) in energy use, energy costs, or energy intensity over the two-to four-year period of the analyses. More specifically, most of the studies reviewed for this report indicate 3 to 8 percent reductions in gross energy consumption or energy use intensity over a two-to four-year period of B\&T policy implementation \cite{mims_evaluation_2017}." Another study found an average savings of 3\% in Austin, New York, San Francisco, and Seattle \cite{palmer_does_2015}. A study on buildings in downtown Chicago showed that their energy disclosure policy was associated with a 6.7\% decrease in vacancy \cite{shang2020impact}. These results have spurred policy changes in hundreds of cities across the world as part of programs such as the C40 Cities\footnote{\url{https://www.c40.org/}} and the Bloomberg American Cities Challenge\footnote{\url{https://www.bloomberg.org/program/environment/climatechallenge/}}.

\subsection{Opening the black box - benefits of explanatory models}
The foundation for benchmarking systems is in machine learning (ML) models that predict how much energy a building would consume based on its attributes and a database of \emph{similar} buildings. The ML community is in a state of reflection on the extent to which various prediction models \emph{can be trusted} \cite{ribeiro_why_2016}. This situation has arisen from experimental studies of ML methods in medical applications. In one study, prediction of pneumonia diagnosis readmission within 30 days was tested for a balance between accuracy and explainability \cite{caruana_intelligible_2015}. The study showed that \emph{explainable} prediction models could achieve high accuracy while proving the trust-building intelligibility that doctors want. An extensive analysis of human subjects and their interactions with the results of machine learning models created a typology of what types of models work best in terms of developing trust in users \cite{madumal_grounded_2019}. 

For building benchmarking applications, various platforms have achieved significant progress. However, there remains some hesitation to use rating systems for decisions beyond the simple transfer of ownership \cite{international_partnership_building_2014}. More considerable energy savings might be possible if decision-makers could understand and trust the underlying model. A recent study found that current energy indicators can misrepresent the retrofit potential in buildings \cite{khayatian_building_2017}. In terms of Energy Star accuracy, a detailed regression study in 2014 showed that of 10 out of 11 building use types, the scores produced had an uncertainty of +/-35\% \cite{scofield_energy_2014}. Another recent review calls attention to model and data accuracy as significant factors to be investigated as well \cite{mims_evaluation_2017}. With explanatory models, analytical tools could be used to find these misrepresentations and improve the confidence a decision-maker has in using the rating system.  

\subsection{Towards more accurate and explainable benchmarking}
In this paper, three questions were addressed: 1) Can the predictive models used in the Energy Star system be improved by using linear or nonlinear models with interaction effects? 2) Using these proposed models, is there an opportunity to improve model interpretability that can enhance the use of benchmarking in the decision-making process? 3) Can these methods be used for an energy benchmarking system using public data sets without using the CBECS data set? If so, what are all the building attributes that significantly influence energy use?

To answer the first question, two approaches were studied: a) linear regression with explicit interaction terms of different orders, b) nonlinear gradient boosting tree models that implicitly use high-level interactions. The predictive performance of these two models is compared with the baseline ordinary regression model, which is used in the Energy Star system and four other models that were used in recent benchmarking studies. Models with interaction terms were shown to offer better accuracy than the status quo models. To answer the second question, it was proposed that nonlinear models with Explainable Artificial Intelligence (XAI) methods be implemented to interpret the model predictions for each building. To answer the third question, public data sources were collected from New York City and Seattle, two cities in the United States. Local publicly accessible energy disclosure policy data sets were combined with tax assessment records for each city. Multiple predictive models were then tested, and they show that public data sets can accurately predict energy use for most building types. 

This paper is organized as follows. In Section~\ref{sec:related_work}, related works were reviewed followed by explaining the data set details in Section~\ref{sec:dataset}. In Section~\ref{sec:methodology}, the methodology was outlined of the proposed predictive models and the methods of explainability. In Section~\ref{sec:implementation}, the details of the experimental implementation were shown, and their accuracy calculated as compared to the conventional Energy Star error rates. In Section~\ref{sec:explainability}, the implementation analysis was extended through the investigation of ways to explain the models for use in building performance analysis. In Section~\ref{sec:discussion}, a critical discussion was provided of the potential impact of such model changes, and explainability could have in a practical sense. Finally, in Section~\ref{sec:conclusion}, an overview of the insights gained was given in addition to discussions of limitations, future work, and reproducibility.

\section{Background and novelty}
\label{sec:related_work}
Buildings are all different in terms of their physical (size, type, geometry, envelope) and operational characteristics (primary function, schedule, occupancy, mechanical, and electrical fixtures). The combination of these heterogeneous factors and weather conditions affect the building energy use in a complex way. When benchmarking a building against its peer group, it is essential to normalize the energy use for all influencing factors in order to enable a \emph{fair benchmarking}. 

There are numerous normalization methods that exist with varying complexity levels. One of the simplest methods \emph{Energy Use Intensity} or \emph{EUI} normalizes a building's energy use for differences in floor area. It is expressed as the total energy use per unit floor area (e.g., $kWh/m^2$). The EUI can also be calculated based on other influential factors and for different types, e.g., energy use per employee in office buildings (e.g., $kWh/employee$). Though EUI is easy to compute and interpret, it fails to account for multiple influential factors, such as physical characteristics, occupancy, and building subsystems, and their combined effects. On the other hand, whole-building energy simulation models, such as EnergyPlus~\cite{crawley2001energyplus}, can quantitatively account for many influencing factors. However, constructing and calibrating these models takes considerable time, effort, and expertise, which limits their \emph{scalability} for a large number of buildings. The efficacy of benchmarking lies in how many factors are used to calculate normalized energy use and their significance.

\subsection{Energy Star methodology}
Energy Star is a widely used peer group benchmarking system for commercial buildings in the USA and Canada. The peer groups are established based on building activity (e.g., office, hospital), using a nationally representative CBECS data set. This survey data set contains detailed building attributes and energy use details from 6720 samples. Energy Star system normalizes building's energy use for differences in building operations by fitting linear regression models between building attributes and energy use. After empirically removing statistically insignificant attributes, the final model for each building type has 5-7 factor variables (See Table ~\ref{tab:energystar_vars}). These linear models are highly interpretable (See Section~\ref{sec:mlr_feature_interactions}). The Energy Efficiency Ratio (EER) of a building is calculated as a ratio between actual source EUI and normalized source EUI, as predicted by the respective peer group model. Thus, a lower EER indicates higher energy performance relative to the peer group. Next, these EER values are translated into a 1-100 percentile ranking by using a Score Lookup Table. This Score Lookup Table is created by using the parameters of a gamma distribution function (shape and scale), which was fit to the cumulative percentage of sorted EER values. A score of 75 or higher is eligible for Energy Star certification, and it can be interpreted as \emph{this building performs better than 75 percent of similar buildings nationwide}.

There are two limitations in the current Energy Star approach, specifically the models used and data set: 1) The underlying weighted linear regression model is inaccurate to model the complex nonlinear relationship between energy use and building attributes; 2) the CBECS data set contains a limited number of buildings (6720) and is infrequently updated (approximately once in every four years). These limitations are addressed by developing more accurate nonlinear and feature interaction models and make them interpretable by using advanced XAI methods (See Section~\ref{sec:mlr_feature_interactions} and Section~\ref{sec:xgboost}). Furthermore, the feasibility of using city-specific public energy disclosure data set is tested to address the second limitation (See Section~\ref{sec:results_nyc_seattle}).

\subsection{Contemporary benchmarking approaches}
Several reviews have covered the diversity of techniques in the building performance benchmarking domain over the years~\cite{ perez-lombard_review_2009, chung_review_2011, li2014methods}. Theses reviews have highlighted the challenges of constructing a fair and scalable energy benchmarking procedure. 

Multiple Linear Regression (MLR) models have been used in many benchmarking studies around the world. These include benchmarking office buildings in Singapore~\cite{lee2008building}, China~\cite{wei2018study} and South Korea~\cite{park2016development,kim2019development}, hotels in Singapore~\cite{xuchao2010benchmarking} and Taiwan~\cite{wang2012study}, and complex campus buildings in China~\cite{ding2018benchmark}. MLR models are easy to implement and interpret due to their linear and additive properties (See Section ~\ref{sec:mlr}), but they fail to model the complex nonlinear relationship between building attributes and energy use, often resulting in poor performance. In response to the limitations of MLR models, some studies have used nonlinear models, such as Decision Trees~\cite{park2016development}, Support Vector Regression~\cite{li2009applying, li2014methods}, and Artificial Neural Networks~\cite{yalcintas2007energy}. Unlike other models, Decision Trees can inherently handle categorical and missing data. In another study, Decision Tree-based models were used to benchmark 1072 office buildings in South Korea~\cite{park2016development}. In comparison with linear models, nonlinear models were proven to achieve better performance~\cite{Wei2018,papadopoulos_grading_2019}. Another issue with MRL models is that they are difficult to interpret, e.g., which factors make a building inefficient. Other contemporary benchmarking studies have also used econometric-based Stochastic Frontier Analysis~\cite{yang2018due,ding2020comparative} and Data Envelopment Analysis~\cite{lee2009benchmarking, yoon2017objective}. These methods attempt to create efficiency frontiers by differentiating error from inefficiency terms~\cite{yang_due-b:_2018}. However, these approaches were found to be sensitive to outliers~\cite{buck_potential_2007}. Another recent study revisited the concept of \emph{heating and cooling degree days} to create a simplified alternative to MLR \cite{guillen2019comparing}.

A few recent studies leveraged the public data sources for building energy benchmarking. New York City’s energy disclosure data set from Local Law 84 (LL84) has been combined with the Primary Land Use Tax Lot Output (PLUTO) data set to develop a novel benchmarking system~\cite{yang_due-b:_2018}. This work is related to parallel work focused on the understanding of building and urban energy use \cite{yang2019due}. Homogeneous peer groups are defined using a CART model followed by stochastic frontier analysis is performed to identifying building energy efficiency levels. In another study, nonlinear gradient boosting machines were used for benchmarking multifamily houses using the New York City’s energy disclosure data set~\cite{papadopoulos_grading_2019}. Both papers showed that using public data sources resulted in more robust results than the conventional models. More recent studies have also used hourly smart meter data for benchmarking buildings based on their load profiles instead of using building attributes. One study used quantile regression to analyze the daily performance of over 500 schools in California and highlighted types of insights not possible in conventional benchmarking systems~\cite{roth_data-driven_2018, roth_benchmarking_2018}. Other recent studies showed the value of outlier detection \cite{ashuri2019data} and discord discovery \cite{park2020good} in the benchmarking and energy analysis process. The most comprehensive recent study compares the use of open data from ten cities to show that using random forest and lasso regression models in benchmarking can outperform CBECS-based models and can pinpoint the essential variables for the prediction process \cite{roth2020examining}.

\subsection{Novelty of proposed approach}

Based on the literature explored for building performance benchmarking, there are still unexplored areas of specific improvement to systems like Energy Star. There have been only preliminary work in finding alternative methods of models for these systems. In addition, the interaction with the \emph{users} of energy benchmarking systems is rarely discussed in terms of how the model can provide clues regarding \emph{how the benchmarking ratings can be used to improve operations decisions}. 

In this study, both linear and nonlinear models were used to improve the prediction accuracy of benchmarking models. Feature interaction terms were explicitly included in MLR models, and this studies the combined effect of building attributes on energy use (See Section~\ref{sec:mlr_feature_interactions}). Though MLR models have been widely in many benchmarking studies, there was limited or no study in the literature that validated the effectiveness of feature interaction models. MLR models with feature interaction terms were shown to improve the model accuracy without compromising their interpretability. Ensemble learning methods were shown for benchmarking by leveraging the recent advancements made by the machine learning community. In particular, the XGBoost~\cite{chen2016xgboost} algorithm was used that offers high accuracy, reduced computing time, and scalability in comparison with other machine learning models. Furthermore, XGBoost does not require a considerable amount of training data, unlike other advanced models, such as deep learning. In this paper, both of these innovative model types were investigated for the first time in the context of both CBECS and city-based energy disclosure data. 

This paper also outlined the first use of the explainability of machine learning models in the context of building performance rating systems. Related work covered in previous work has worked towards the updates in the accuracy of models for benchmarking, but this paper outlined the first use of SHapley Additive exPlanation or SHAP values~\cite{lundberg2017unified} and visualizations (Section \ref{sec:shap}) for the purpose interpretability and explainability of models. Finally, the proposed approach was evaluated and compared with the Energy Star system and four recent approaches by benchmarking a large number of buildings (approximately 15,000) with six building types.

\section{Open source data sets}
\label{sec:dataset}

One of the significant barriers of energy benchmarking systems is establishing a peer group of buildings with similar characteristics. This section describes the publicly available data that was collated from various sources. These data are at the forefront of prototyping techniques for the purpose of building performance analysis and benchmarking at city-scale levels.

\subsection{CBECS data set}
The Energy Star benchmarking system relies on the Energy Information Administration’s (EIA) CBECS data set, which establishes a peer group for each building type separately. Energy Star also relies on other survey data sets for a few building types such as hospitals, multifamily housing, senior care communities, and wastewater treatment plants. More details are given in their technical documentation~\footnote{\url{www.energystar.gov/ENERGYSTARScore}}. 

The CBECS data set contains detailed building characteristics for 6,720 buildings across several cities within different climates to represent the USA building stock. More details about the questionnaire and sampling method followed for selecting the buildings are described in the EIA website. There is a variable in the data set called \emph{FINALWT} that represents the final full sample weight for each building. This weight attribute is utilized in the Energy Star system for developing predictive models. The CBECS data set is leveraged first to reproduce the predictive models used in the Energy Star system. Second, alternate models are also investigated using this data set, and the same set of variables as used in the Energy Star system. By running both sets of models, one can empirically validate the model performance of Energy Star’s system against alternative models. 

\subsection{Seattle and New York City data sets}

Over 20 cities across the world have started to implement energy disclosure policies with the goal of reducing energy demand from buildings. These policies require buildings with specific minimum gross floor area to collect and share their energy usage details along with some specific building attributes for benchmarking. Cities then mandate the benchmarking results, as well as other building-specific information like EUI, to be released for public use. These data sets typically contain information about each building’s address, primary use type, age, yearly total energy consumption, total gross floor area, EUI, and Energy Star score. To supplement the information in these data sets, data from publicly available tax assessor databases were collected that contain additional information regarding building attributes. The property tax assessment records contain additional information about the buildings, such as the number of floors, number of units, geometrical attributes, and some electrical load detail. These two data sets were merged by using either unique building identification codes or geocoding the data set and merging on latitude and longitude coordinates. After merging, extensive cleaning was completed to remove duplicates and inconsistent building attributes. In this work, public data sets for two cities in the USA cities of New York and Seattle were collected. These two cities were selected because of the availability of a large number of building characteristics.

\begin{table}[t!]
    \caption{List of building types and number of samples in the CBECS, New York City and Seattle open data sets.} 
    \label{tab:dataset_summary}
    \centering
    \small
    \begin{tabular}{lrrr}
        \toprule
        Building type & CBECS & NYC & Seattle \\ 
        \midrule
        Hotels & 104 & 195 & 70 \\ 
        K-12 Schools & 333 & 106 & 122 \\ 
        Multifamily housing & 319 & 8,206 & 1,270 \\ 
        Offices & 628 & 1,609 & 429 \\ 
        Retail stores and supermarkets & 179 & 989 & 113 \\ 
        Worship facilities & 249 & 26 & 69 \\ 
        \midrule
        Total & 1,812 & 11,131 & 2,073 \\ 
        \bottomrule
    \end{tabular}
\end{table}

\subsection{Data cleaning and preprocessing}
In the Energy Star system, a series of filters were applied over the CBECS data set, namely building type, program, data limitation, and analytical filters, to make a nationally representative and homogeneous peer group. In order to make a fair comparison of the predictive models, the same filters were applied as mentioned in the respective technical document for each building use type\footnote{There are some minor differences in the total number of buildings mentioned in their technical documentation and the proposed implementation. The differences arise primarily due to some unclear details in the document. More details can be found in the code repository for this work}. 

For the Seattle data set, buildings from energy benchmarking reports and property tax records were matched based on the unique tax parcel identification number, which was present in both the data sources. Whereas, for New York City, the ten digits identifying the unique Borough Block and Lot (BBL) number were used to merge the two data sources. 

Since building attributes were collected from public sources, there were many outlier samples. In Energy Star, outliers are removed by applying various data limitations and analytical filters.  For example, there is an upper limit on the gross floor area (1 million for most building types) and source EUI for most of the building types. All buildings were removed with source EUI less than one percentile and more than 99 percentile in the sample set.

After cleaning the data, the buildings were grouped with similar activities, as done in the Energy Star system. For example, office, bank/financial institutions, and courthouses were collectively referred to as office buildings and benchmarked as a group. In order to make a fair comparison, the same definitions were followed for grouping similar buildings based on their primary property type, which was given in the energy benchmarking data sets. 

Table \ref{tab:dataset_summary} shows the list of building use types and their corresponding number of buildings from each data set after cleaning.  Only these six building use types were selected because of the availability of predictive model details, as per the Energy Star technical documentation, and public data sets. Moreover, these six use types represent the majority of the buildings in both the cites (94\% in New York and 82\% in Seattle). In the Energy Star system, a different number of independent variables were used as predictors for modeling source EUI. Unlike Energy Star, all the available relevant building attributes were used as features for fitting the models.

\section{Methodology}
\label{sec:methodology}

Machine learning algorithms are widely used for modeling the energy performance of buildings~\cite{Wei2018}. They are broadly classified as \emph{linear} and \emph{nonlinear} on the basis of how they model the relationship between the dependent and independent variables. Linear models are expressed as a weighted sum of independent variables and an error term. The learning task involves computing weights (parameters or coefficients) of each independent variable from training samples. Linear models are simple, intuitive, and easy to interpret. However, they are inadequate for modeling complex relationships that are inherent to many real-world systems. As a result, they are often found to be inaccurate in many energy management applications~\cite{Wei2018,wang2018review}. 

Nonlinear models can handle complex relationships between dependent and independent variables, and they outperform linear models in terms of accuracy in many application domains~\cite{Wei2018,wang2018review}. However, nonlinear models lack interpretability as it is difficult to understand the influence of a single independent variable. Independent variables are convolved in a complex manner to predict the dependent variable. Selecting accurate and interpretable models is essential for developing an excellent benchmarking system that can rank buildings and reveal the causes of energy efficiency or inefficiency.

In this work, both linear and nonlinear models were used. Among the two predictive models employed, the first one was the widely used MLR. Feature interaction terms were included, that slightly relaxed the linearity assumption, to MLR model for improving accuracy while retaining their interpretability (Section~\ref{sec:mlr}). The second model was the XGBoost~\cite{chen2016xgboost}, a nonlinear decision tree-based ensemble learning method (Section~\ref{sec:xgboost}). XGBoost models were augmented with SHAP values~\cite{lundberg2017unified} for making the model predictions interpretable (Section~\ref{sec:shap}). These approaches helped in developing a more accurate and explanatory building energy benchmarking system.

\subsection{Multiple linear regression}
\label{sec:mlr}
MLR is a widely used technique for modeling the linear relationship between a dependent (outcome) variable and one or more independent variables (predictors). It is formulated as:

\begin{equation}
Y_i = \beta_0 + \sum_{j=1}^{p} \beta_j X_{i,j} + \epsilon_i, \quad i = 1, 2, ..., n
\end{equation}

Here $n$ is the number of samples, $p$ is the number of predictors, $Y_i$ is the outcome variable, $X_{i,j}$ is a vector of $p$ predictors, $\epsilon_i$ is the error term for the $i$th sample, and $\beta_0$ is the offset term. The weights $\beta_*$ are usually calculated using Ordinary Least Squares (OLS)~\cite{friedman2001elements}. 

The two most essential attributes of the MLR model are \emph{additivity} and \emph{linearity}. The additive property states that the relationship between each predictor variable and outcome is independent of other predictors. The effect of a change in the value of one variable on the outcome is independent of changes in other variables. The linearity property (constant weights $\beta_j$ for each predictor) states that there is a constant change in the outcome for a unit change in the predictor variable, regardless of the value of the variables. These two properties make MLR interpretable because the effect of each predictor on the outcome is separable. However, linearity and additive may hot hold in many real-world applications. As a result, MLR models often found to be inaccurately mapping the relationship between outcome and predictors.

\subsection{Multiple linear regression with feature interactions}
\label{sec:mlr_feature_interactions}
When two are more independent variables are involved in a model, they may conspire to affect the dependent variable jointly. This is broadly known as \emph{synergy} or \emph{interaction}~\cite{friedman2001elements}~\cite{aiken1991multiple}. The presence of interactions indicates that the effects of independent variables are compounded. Interactions are inherent to many natural phenomena, and they have been extensively studied in many disciplines, such as life sciences~\cite{cordell2009detecting,braha2003use}. The advantage of adding interaction terms to a linear model is two-fold. First, it unveils the presence of unknown interaction effects, if any, that influence the dependent variable, which is otherwise difficult to identify using experiments. Second, it helps to increase the model accuracy as the inclusion of additional predictors, representing the variable interactions, better model the relationship among the variables of interest.

This work was founded on the hypothesis that \emph{there are interactions among building attributes that influence the energy usage of a building, beyond their individual effects. Hence, the inclusion of interaction terms would significantly improve the accuracy of MLR model}. The inclusion of these interaction terms would further help in identifying and quantifying the interactions among building attributes if any. While the independent effect of each predictor (building attributes) was quantified on energy use using the MLR model, it was essential to quantify the effect of the interaction terms. 

A MLR model with second-order interaction terms can be written as:
\begin{equation}
Y_i = \beta_0 + \sum_{j=1}^{p} \beta_j X_{i,j} + \sum_{j>k}^{} \beta_{j,k} X_{i,j}X_{i,k} + \epsilon_i, \quad i = 1, 2, ..., n
\end{equation}

The terms $X_{i,j}X_{i,k}$ denote the pairwise interaction between two predictors $X_{*,j}$ and $X_{*,k}$. With $p$ predictors, there are $\binom n2$ second-order interaction terms that can be included in an MLR model. Based on the rules of hierarchy, it is required to include main effects and all low-order interaction terms when higher-order interaction terms are included in an MLR model~\cite{friedman_elements_2001}. Thus, the maximum number of additional features that can be included in a $m$-order interaction model is $p + \sum_{j=2}^{m} \binom pj$. Though all possible higher-order interaction terms to a model can be included, not all interactions would be statistically significant. The final model will contain only the most significant interaction terms, together with their main effects.


\subsubsection{Interpreting MLR models with feature interactions}
\label{sec:mlr_interpretation}
Interpretation of the MLR model with interaction terms was similar to the ordinary MLR model, although additional conditions were necessary~\cite{aiken1991multiple,friedman2001elements}. For example, consider the regression with two independent variables, Gross Floor Area (GFA)  and occupancy percentage (OCC), for modeling EUI as:

\begin{equation}
EUI = b0 + b1 \cdot GFA + b2 \cdot OCC + \epsilon
\end{equation}

Here $b0$ is the offset (baseline EUI), $b1$ and $b2$ are the coefficients for the predictors $GFA$ and $OCC$, respectively. This simple model can be interpreted as: if we increase $GFA$ by one unit, then $EUI$ will increase by an average of $b1$ units when $OCC$ is fixed, and vice versa. Including an additional predictor representing the interaction of $GFA$ and $OCC$, the regression equation becomes
\begin{equation}
EUI = b0 + b1 \cdot GFA + b2 \cdot OCC + b3 \cdot (GFA \cdot OCC) + \epsilon
\end{equation}
This equation can be rewritten as 
\begin{equation}
EUI = b0 + b1 \cdot GFA + (b2 + b3 \cdot GFA) \cdot OCC  + \epsilon
\end{equation}
The weight $b3$, the coefficient of the interaction term, can be interpreted as 
\emph{the increase in the effect of $GFA$ on $EUI$ for one unit increase in $OCC$, or vice versa}. 
Note that as the model complexity increases by including higher-order interaction terms, it becomes less intuitive to interpret. 


\begin{center}
\begin{figure*}[t!]
    \centering
    \includegraphics[scale=0.6]{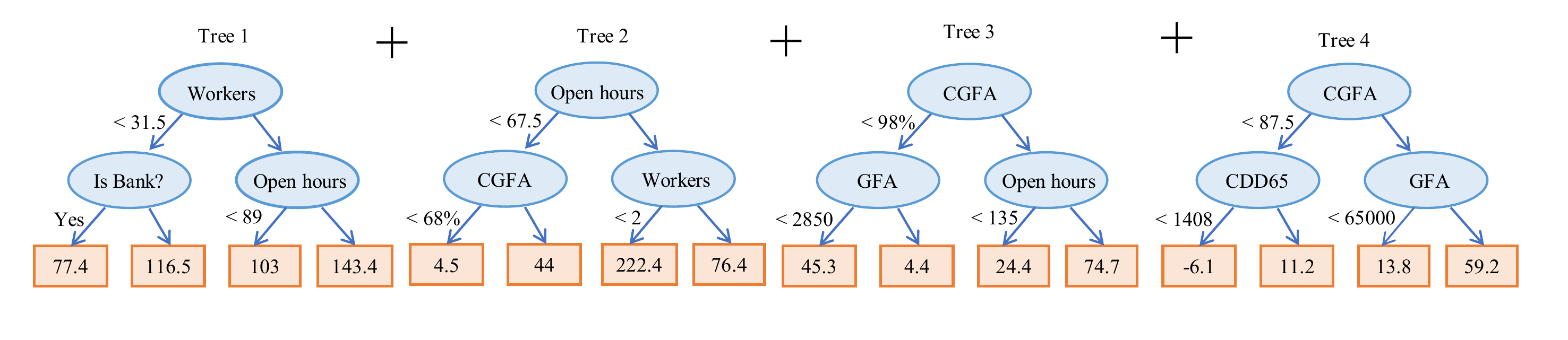}
    \caption{An example XGBoost model for predicting EUI using six building attributes (See Table~\ref{tab:office_vars}). This simple model consisted of four classification and regression trees (CART), each of height two. The leaves contained the prediction score. The model was constructed by sequentially adding trained trees together in a greedy manner. Subsequent trees were trained on sub-samples that were designated as harder to predict by the preceding trees. The final prediction score was the sum of prediction from each tree.}
    \label{fig:gbt}
\end{figure*}

\end{center}

\subsection{Gradient boosted trees - XGBoost}
\label{sec:xgboost}

Gradient Boosted Trees (GBT) have attracted the attention of the machine learning community in recent years~\cite{friedman2001greedy}. GBTs fall under the broader \emph{ensemble learning} category in which many base models are combined to make a single better prediction model. An optimized and distributed implementation was used of GBT called the eXtreme Gradient Boosting (XGBoost) library~\cite{chen2016xgboost}. XGBoost offered high accuracy, reduced computing time, and scalability. 

An XGBoost model consists of a set of Classification and Regression Trees (CART)~\cite{breiman1984classification}. The structure of a CART model is similar to a \emph{binary search tree} in which every node has two children except leaves. Data samples are partitioned into two groups at each node in a hierarchical manner until no further split is possible to create leaf nodes with target values. The partitioning is usually based on \emph{information gain} of the samples. Due to the way it splits the data into two subgroups at each level and its hierarchical structure, CART models are highly interpretable. One can travel along with the nodes from the root to the leaves to explain how the model makes predictions. 

In XGBoost, CARTs are used as base learners, and they are combined using a \emph{tree boosting} technique~\cite{chen2016xgboost}. The model is constructed by sequentially adding trained CART models together. Subsequent trees are trained on sub-samples that were hard to predict by the preceding trees.
The final score of the model is calculated by summing up the prediction scores of each individual CART model. The detailed description of the algorithm can be found in~\cite{chen2016xgboost}. An example XGBoost model for predicting EUI using six building attributes is shown in Figure~\ref{fig:gbt}.

\subsubsection{Tuning hyper-parameters and model selection}
\label{sec:xgboost_tuning}
XGBoost offered several parameters that could be tuned to select an optimal model. A grid search method was employed for tuning four model parameters. The list of hyper-parameters and their grid search ranges are given in Table~\ref{tab:gbt_params}. A 10-fold cross-validation method was used with two repeated rounds using root mean squared error as the performance metric to avoid model over-fitting. In each cross-validation iteration, the entire data set was split into training (90\%) and testing (10\%) sets, and a model was fitted using a hyper-parameter combination. This process was repeated ten times using different combinations of training and testing set, and the error from each iteration was averaged. A model with the lowest error rate was selected as the final model.

\begin{table}[t!]
    \caption{List of hyper-parameters tuned and their grid-search range for GBT.} 
    \label{tab:gbt_params}
    \centering
    \small
    \begin{tabular}{llr}
        \toprule
        Parameter & Description & Search range \\ 
        \midrule
        max\_depth & Max Tree Depth & 2-3 \\ 
        nrounds & \# Boosting Iterations & 1-200 \\ 
        eta & Shrinkage & 0.1-0.9 \\ 
        colsample\_bytree & Subsample Ratio of Columns & 0.2-0.8 \\ 
        subsample & Subsample Percentage & 0.25-1 \\ 
        \bottomrule
    \end{tabular}
\end{table}

\subsection{Feature interactions in XGBoost models}
\label{sec:xgboost_interactions}
The hierarchical structure of decision trees makes them a powerful tool in automatically capturing high-level variable interactions. The decision variables that appear together in the path from the root to each leaf node are said to be interacting with each other~\cite{breiman1984classification}. The height of the tree is equivalent to the highest order of interactions that can be included in the model. For example, there were two second-order interactions that existed in each tree in the XGBoost model, as shown in Figure~\ref{fig:gbt}. Also, there were three interaction terms ($WorkersCnt \cdot OpenHours$, $OpenHours \cdot CGFA$, and $CGFA \cdot GFA$), out of a total of eight, that appeared twice in the model. This appearance indicated that they were the most influential predictors as compared to the remaining five. In this work, the performance of XGBoost models was analyzed with different orders of interactions by limiting the tree height (See Section~\ref{sec:results_cbecs}). The interactions that exist between building attributes were studied as well as their level of influence in predicting the energy use by systematically measuring their strength using SHAP values (See Section~\ref{sec:results:xgboost_shap}).

\subsection{Interpretation of XGBoost models using SHAP values}
\label{sec:shap}
As an ensemble algorithm, it is difficult to explain the predictions of XGBoost models out of the box. To mitigate this problem, this work proposes XGBoost with Explainable Artificial Intelligence (XAI) methods that are capable of unboxing black-box models. Unlike the classic feature importance measures that focused on whole model interpretation (global), these modern XAI methods enable interpretations of even individual predictions of the model (local).

\emph{SHapley Additive exPlanation values}, or \emph{SHAP values}, were used for interpretation~\cite{lundberg2017unified}. This metric belongs to the class of \emph{additive feature attribution methods} in which a model's prediction is explained as a sum of values attributed to each feature. An explanation model $g$, representing the interpretable approximation of the original model, is defined as a linear function of binary variables:

\begin{equation}
g(z') = \phi_0 + \sum_{i=1}^M \phi_iz_i'
\end{equation}

Here, $z' \in \{0,1\}^M$ denotes the presence or absence of a feature, $M$ is the number of features in the model, and $\phi_i \in \mathbb{R}$ are the feature attribution values. The SHAP values, that attribute $\phi_i$ to each feature, are estimated by combining the conditional expectation $E[f(x) | x_S]$ of all subset of features with Shapley values~\cite{shapley1953value}:

\begin{equation}
\phi_i = \sum_{S \subseteq N \setminus\{i\}} \frac{ |S|!\left(M-|S|-1\right)!} {M!} [f_x \left(S\cup\{i\}\right) - f_x\left(S\right)]
\end{equation}

Here, $N$ is the set of all features, $S$ is a subset of $N$ with non-zero indexes in $z'$, and $f_x(S) = E[f(x) | x_S]$ is the expected outcome of the model conditioned on $S$. SHAP values are model-agnostic. The computation time of SHAP values increases exponentially with the number of features in the model. Previous literature has proposed a novel polynomial-time algorithm specifically for tree ensemble models such as XGBoost~\cite{lundberg2018consistent}.

An extension of SHAP values called \emph{SHAP interaction values}~\cite{lundberg2018consistent}, for measuring the second-order interactions between feature $i$ and $j$, denoted as $\Phi_{i,j}$, is defined as:

\begin{equation}
\Phi_{i,j} = \sum_{S \subseteq N \setminus\{i,j\}} \frac{ |S|!\left(M-|S|-2\right)!} {2\left(M-1\right)!} \nabla_{ij}\left(S\right)
\end{equation}

Here, $\nabla_{ij}\left(S\right) = f_x(S\cup\{i,j\}) - f_x(S\cup\{i\}) - f_x(S\cup\{j\}) + f_x(S)$ when $i \neq j$. Note that $\Phi_{i,j} = \Phi_{j,i}$ and the total interaction effect is $\Phi_{i,j} + \Phi_{j,i}$. 
$\nabla_{ij}$ denotes the pure interaction between feature $i$ and $j$, after accounting for main and their individual effects from total combined effects. $\nabla_{ij}$ is averaged over all feature coalitions $S$ to get the final pairwise interaction matrix with dimensions $M \times M$. Furthermore, the main effects of feature $i$, when $i = j$, can be defined as the difference between SHAP and SHAP interaction values:

\begin{equation}
\Phi_{i,i} = \Phi_i - \sum_{j \neq i} \Phi_{i,j}
\end{equation}

SHAP values are the only unique measure with three desirable properties: \emph{local accuracy}, \emph{missingness}, and \emph{consistency}~\cite{lundberg2017unified}. A python-based SHAP library was used for the calculation and visualization of SHAP values of XGBoost models\footnote{\url{https://github.com/slundberg/shap}} (See Section~\ref{sec:results:xgb_interpretation}).

\section{Model implementation}
\label{sec:implementation}

The first objective in this comparative analysis was to implement the two proposed modeling techniques to assess their performance relative to the MLR and recent studies based on standard error metrics. These error metrics were defined to outline the model implementations on the CBECS data set as well as on two open data sets from New York and Seattle.

The implementation included predictive models for six building types. Source energy ($kBtu$) was used as the model output, and all variables listed in the Energy Star technical documentation for the respective building types were used as predictors. The list of predictor variables for different building types is shown in Table~\ref{tab:energystar_vars}. Only these variables were used to show how much the increase in accuracy can be achieved by using models with feature interactions. Whereas on the public data sets from New York City and Seattle, total energy was used as a target variable, and all available building attributes were used as predictors. 

\begin{table}[ht]
    \centering
    \footnotesize
    \begin{tabular}{ll}
        \toprule
        Variable &  Description \\ 
        \midrule
        GFA &  Gross Floor Area (m\textsuperscript{2}) \\      
        CGFA &  Cooled Gross Floor Area (\%) \\     
        WorkersCnt &  Number of employees (per 92.9 m\textsuperscript{2} (1000 ft\textsuperscript{2})) \\ 
        ComputersCnt &  Number of computers (per 92.9 m\textsuperscript{2} (1000 ft\textsuperscript{2})) \\ 
        OpenHours &  Total hours open per week \\ 
        CDD65 &  Cooling Degree Days (base 18.3\ensuremath{^\circ}C (65\ensuremath{^\circ}F)) \\       
        IsBank & Whether or not a bank branch (yes=1 and no=0) \\       
        \bottomrule
    \end{tabular}
    \caption{List of building attributes used in the office building model.} 
    \label{tab:office_vars}
\end{table}

\begin{table}[ht]
    \centering
    \begin{tabular}{l>{\centering}*{2}{c}}
    \toprule
    Building attribute & New York City & Seattle \\ 
    \midrule
    Gross Floor Area (m\textsuperscript{2}) & \checkmark & \checkmark \\
    Age of the building & \checkmark & \checkmark \\
    Number of buildings & \checkmark & \checkmark \\
    Number of floors & \checkmark & \checkmark \\
    Average occupancy (\%) & \checkmark &  \\
    Number of residential units & \checkmark &  \\
    Number of total units & \checkmark &  \\
    Building quality  &  & \checkmark \\
    Construction class &  & \checkmark \\
    Heating system type &  & \checkmark \\
    Building shape &  & \checkmark \\
    \bottomrule
    \end{tabular}
    \caption{List of building attributes in the New York City and Seattle public data sets after cleaning.} 
    \label{tab:nyc_seattle_vars}
\end{table}


\subsection{Performance metrics}
\label{sec:metrics}
Three metrics were used for comparing the predictive performance of models described in Section~\ref{sec:methodology}. 

\begin{itemize}

\item{$Adjusted~R^2$:}
The coefficient of determination or $R^2$ is a widely used \emph{relative measure} in regression analysis. 
It is defined as: 

\begin{equation}
R^2 = 1 - \frac{\sum_{i=1}^{n} (y_i- \hat{y}_i)^2}{\sum_{i=1}^{n} (y_i- \bar{y})^2}
\end{equation}

Here $n$ is the number of observations, $y_i$ and $\hat{y}_i$ are the observed and predicted values, respectively, of the $i^{th}$ observation and $\bar{y}$ is the mean of observed values.

The Energy Star system uses $R^2$ to quantify the explanatory power of regression models. However, $R^2$ is not a fair measure for comparing models with different numbers of predictors, and this would increase when more variables are included in the model. To address this issue, a robust measure called \emph{adjusted $R^2$} was adopted. The adjusted $R^2$ increases only if the added term statistically improves the model.
The adjusted $R^2$ is defined as:
\begin{equation}
\bar{R}^2 = 1 - (1-R^2) \frac{n-1}{n-p-1}
\end{equation}
Here $p$ is the number of independent variables in the model. Please note that adjusted $R^2$ is weighted by sample weights, if available, in the data set.


\item{\emph{Root Mean Squared Error (RMSE)}:} This is the standard deviation of unexplained variance in the model. In contrast with adjusted $R^2$, $RMSE$ is an \emph{absolute measure} which is expressed in the same units as the dependent variable. It is defined as:
\begin{equation}
\text {RMSE} = \sqrt{ \frac{1}{n} \sum_{i=1}^{n} (y_i- \hat{y}_i)^2 }
\end{equation}

\item{\emph{Normalized Root Mean Squared Error (NRMSE)}:} This is the normalized RMSE that facilitates comparison of models with different scales, e.g. building types. RMSE was normalized by the range of the dependent variable and expressed as a percentage. $NRMSE$ is defined as:
\begin{equation}
\text {NRMSE} = \frac{RMSE}{y_{max} - y_{min}} \times 100\%
\end{equation}

\item{Mean Absolute Percent Error (MAPE):} We used MAPE to report the prediction error as percentage. This metric was used for comparing model performance between different public data sets. It is defined as:
\begin{equation}
\text {MAPE} = \frac{100\%}{n}  \sum_{i=1}^{n}  \abs{\frac{y_i-\hat{y}_i}{y_i}}
\end{equation}

\end{itemize}

\begin{figure}[t!]
	\centering
	\includegraphics[scale=1]{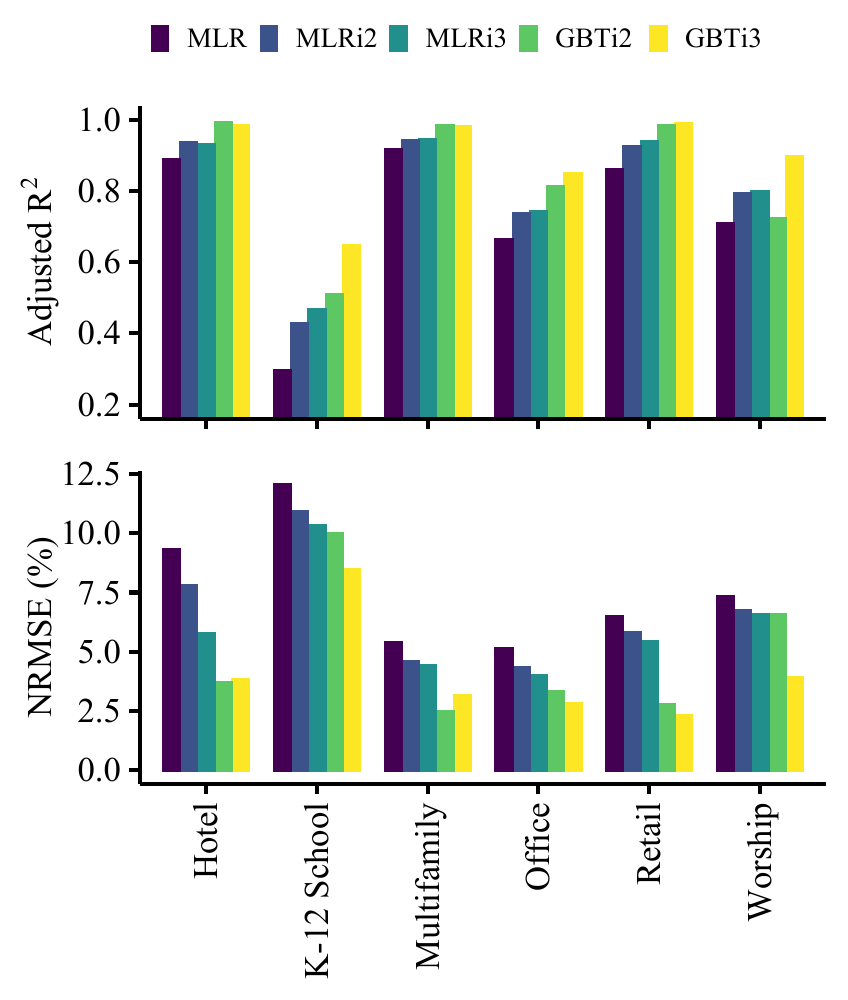}    
	\caption{Comparison of NRMSE between different predictive models for six building types in the CBECS data set. MLR was the baseline model which was used in the Energy Star system. All MLR and GBT models with second and third order interaction terms achieved lower NRMSE for all six building types.}
	\label{plot:cbecs_r2_nrmse}
	\vspace{-6pt}
\end{figure}

\begin{figure}
	\centering
	\includegraphics[scale=0.7]{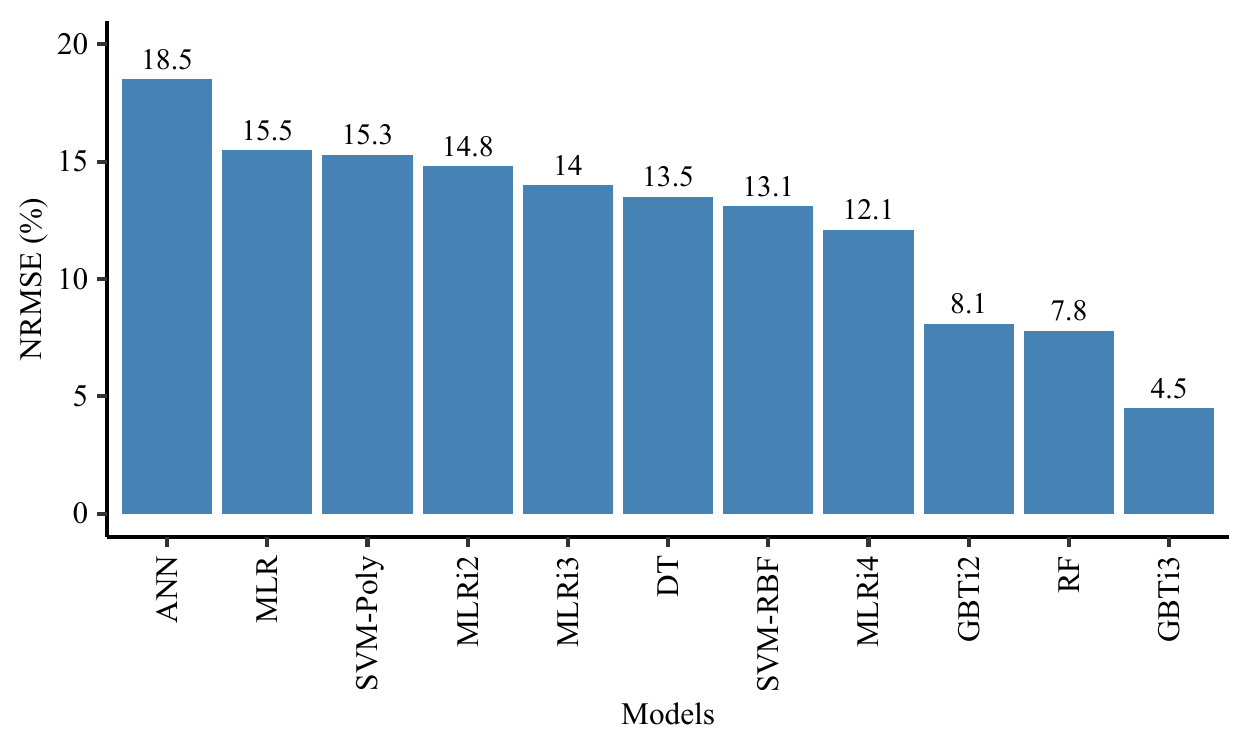}
	\caption{Comparison of average NRMSE between the proposed (linear: MLRi2, MLRi3, and MLRi4; nonlinear: GBTi2 and GBTi3) and the baselines models across six buildings. The nonlinear GBTi3 model achieved the lowest error as compared to all other models. The linear MLRi4 model also achieved the lowest error with the exception of the nonlinear Random Forest (RF) model.}
	\label{fig:cbecs_nrmse_all}
\end{figure}

\begin{table}[t!]
	\centering
	\small
	\begin{tabular}{lrrrr}
		\toprule
		Metrics & MLRi2 & MLRi3 & GBTi2 & GBTi3  \\ 
		\midrule
		Increase in adjusted $R^2$ & 2.3 & 4.9 & 13.8 & 24.9 \\
		Decrease in NRMSE & 3.2 & 7.0 & 9.1 & 13.7 \\
		\bottomrule
	\end{tabular}
	\caption{The average increase in adjusted $R^2$($\%$) and average decrease in NRMSE across all six building types when compared to Energy Star's ordinary MLR models with the proposed linear interaction models (MLRi2 and MLRi3) and nonlinear models (GBTi2 and GBTi3).} 
	\label{tab:cbecs_results_eui}	
\end{table}

\subsection{CBECS data set}
\label{sec:results_cbecs}
The performance of the ordinary weighted MLR, which is used in Energy Star, was compared with four models that have been described in the previous section --- two MLR and two GBT models using second and third-order interaction terms. These proposed models were labeled as MLRi2, MLRi3, GBTi2, and GBTi3. Only MLR models were selected with up to third-order interaction terms as the inclusion of additional interactions would lead to overfitting. In all the interaction models, the total number of predictors was bounded by $1/3$ of the number of samples in the data set. This approach is common practice in the machine learning community.

The comparison of adjusted $R^2$ and NRMSE values of all predictive models for different building types using source EUI as the dependent variable is shown in Figure~\ref{plot:cbecs_r2_nrmse}. Further, the average increase in adjusted $R^2$ and the average decrease in NRMSE across all six building types when comparing the ordinary MLR model with interaction models are summarized in Table~\ref{tab:cbecs_results_eui}. From Figure~\ref{plot:cbecs_r2_nrmse} and Table~\ref{tab:cbecs_results_eui}, it can be observed that all interaction models achieve higher adjusted $R^2$ and lower NRMSE than ordinary MLR models. The MLR model with second-order interaction terms performs better than the ordinary MLR model for all six building types. MLRi2 achieves 2.3\% increase in adjusted $R^2$ and 3.2\% decrease in NRMSE on average than the MLR model. Between MLRi2 and MLRi3, MLRi3 performs better for all building types both in terms of increase in adjusted $R^2$, except for worship facilities, and decrease in NRMSE. These results reveal that there were significant interactions exist among building attributes.

\begin{figure*}[ht!]
	\centering
	\scriptsize
	\includegraphics[scale=0.7]{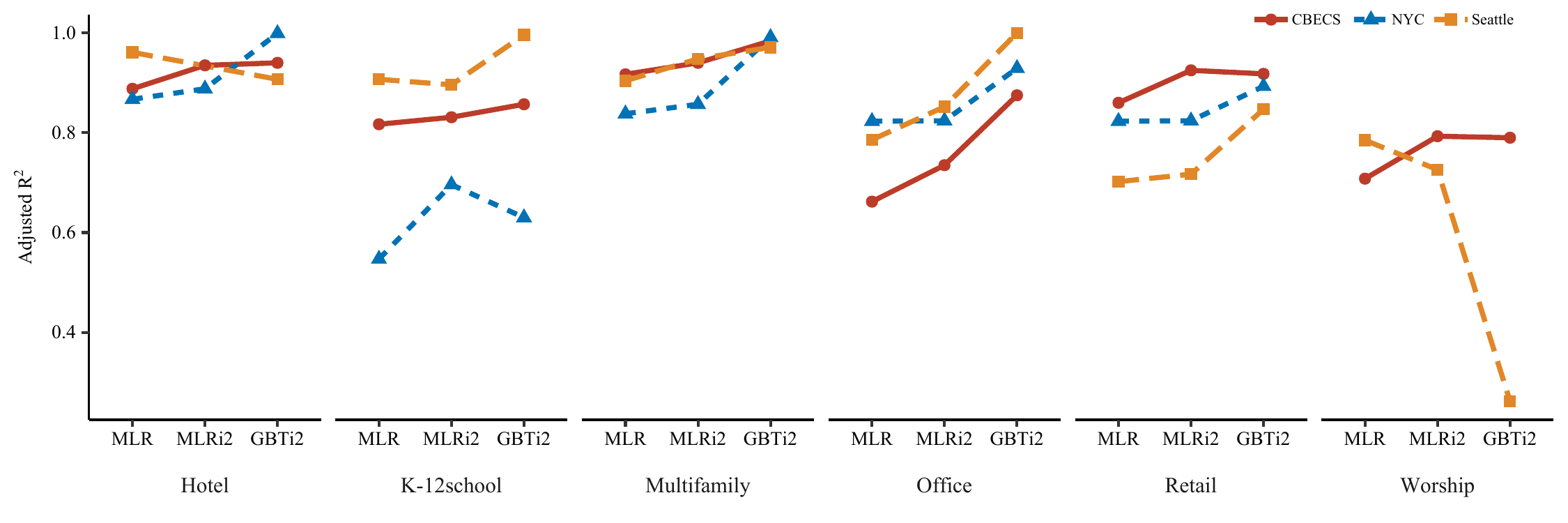}
	\vspace{-10pt}
	\caption{Comparison of adjusted $R^2$ values of three models for six building types on CBECS, New York City, and Seattle data sets.}    
	\label{fig:all_r2}
\end{figure*}

\begin{figure*}[ht!]
	\centering
	\scriptsize
	\includegraphics[scale=0.7]{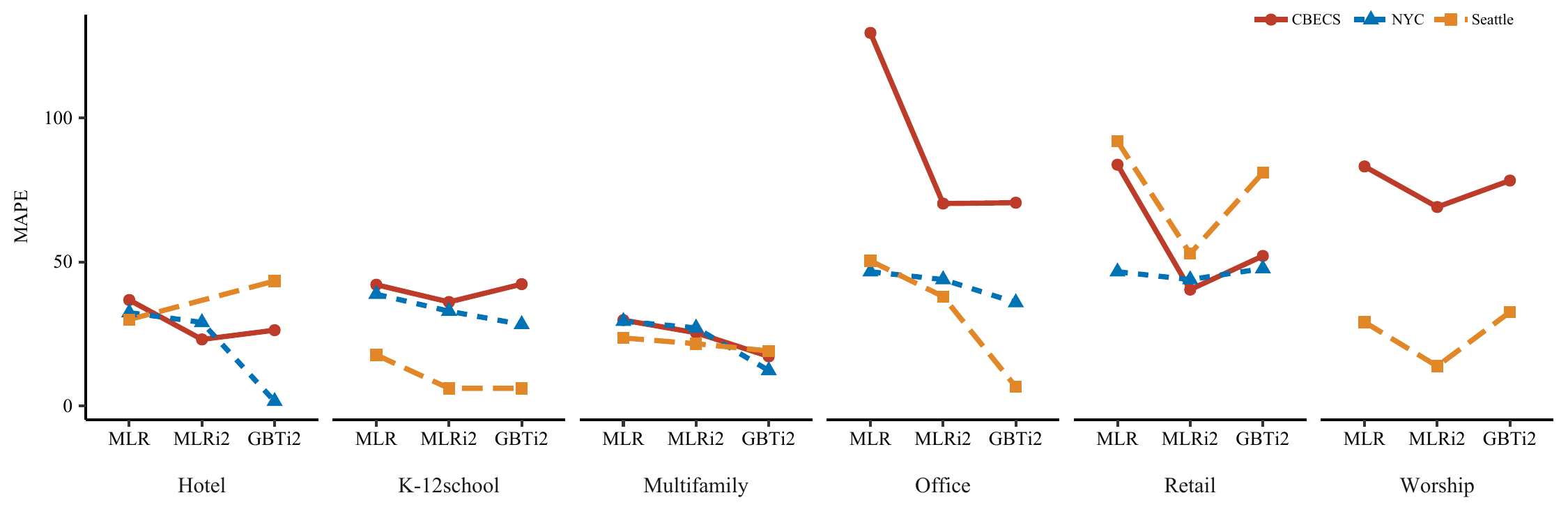}
	\vspace{-10pt}
	\caption{Comparison of MAPE values of three models for different predictive models for six building types on CBECS, New York City, and Seattle data sets.}
	\label{fig:all_mape}
\end{figure*}

Overall, both the nonlinear GBT models perform better than all linear MLR, ordinary, and interaction models. GBTi2 and GBTi3 models achieved a 13.8\% and 24.9\% increase in adjusted $R^2$ and a 9.1\% and 13.7\% decrease in NRMSE, respectively, on average as compared the baseline MLR model. Between linear MLRi2 and nonlinear GBTi2 models, both with fixed second-order interaction terms, GBTi2 performs as good as or better than MLRi2 for all six building types. Furthermore, though the overall performance of GBTi2 was better than MLRi3, MLRi3 performs slightly better than GBTi2, in terms of adjusted $R^2$, for K-12 schools and office buildings. However, the GBTi3 model, with fixed three-order interaction terms, performs better than all other models in terms of both increases in adjusted $R^2$ and a decrease in NRMSE on average. Overall, GBT performs better because it is a nonlinear model that captures interactions between building attributes and energy usage.

Moreover, the effectiveness of the proposed models was compared with four machine learning algorithms that were used in recent energy prediction studies. These include, a) Decision Trees~\cite{park2016development}; b) Artificial Neural Networks~\cite{yalcintas2007energy}; c) Support Vector Regression (SVR)~\cite{li2009applying, li2014methods} using two different kernels (polynomial and radial basis function); and d) Random Forest~\cite{breiman2001random}. A comparison of NRMSE between the linear (MLRi2, MLRi3, and MLRi4) and nonlinear (GBTi2 and GBTi3) models, and the baseline MLR and recent models, is shown in Figure~\ref{fig:cbecs_nrmse_all}. It can be observed that the proposed GBTi3 model achieves the lowest error (4.5\%) among all other models. Moreover, the linear MLRi4 model with 4-way interaction terms also achieves lower error (12.1\%) than the nonlinear ANN and SVM-based models. This proves that MLR models with high-order feature interaction terms were better than nonlinear models except for the ensemble models. ANN achieves the highest error due to the unavailability of large samples in the data sets.

\subsection{New York and Seattle public data sets}
\label{sec:results_nyc_seattle}
The predictive performance of proposed interaction models was analyzed on two public data sets -- New York City and Seattle. In contrast with using specific attributes for each building type, as is the case in the Energy Star system, all relevant building attributes available were used in the data sets as the predictors of energy use. The predictive performance of the ordinary MLR, MLR, and GBT with 2-order interaction terms was compared using adjusted $R^2$ and MAPE values. While adjusted $R^2$ values show the variance explainable by all predictors, MAPE shows the error ratios. MAPE was specifically used because the total energy consumption varies across the three data sets.

Figure~\ref{fig:all_r2} shows the comparison of adjusted $R^2$ values between three models for six different building use types on CBECS, New York, and Seattle data sets. It shows that for all different building types, except retail and worship, the adjusted $R^2$ values were equal or higher when using NYC and Seattle data sets than CBECS. For retail buildings, all three models achieved lower $R^2$ values. As a result, the CBECS data set was better suited for selecting reference retail buildings than either of the city-specific data sets. For worship facilities, the NYC data set was ignored as it contained very few instances. Both MLRi2 and GBTi2 models perform poorly on Seattle's worship buildings as compared with the ordinary MLR model. It can be concluded that the public building attributes available in the NYC and Seattle data sets were more accurate at modeling the variance in energy use better than CBECS data set for hotels, K-12 schools, multifamily housings, and offices. 

\begin{figure*}[ht!]
	\centering
	\small
	\includegraphics[scale=0.6]{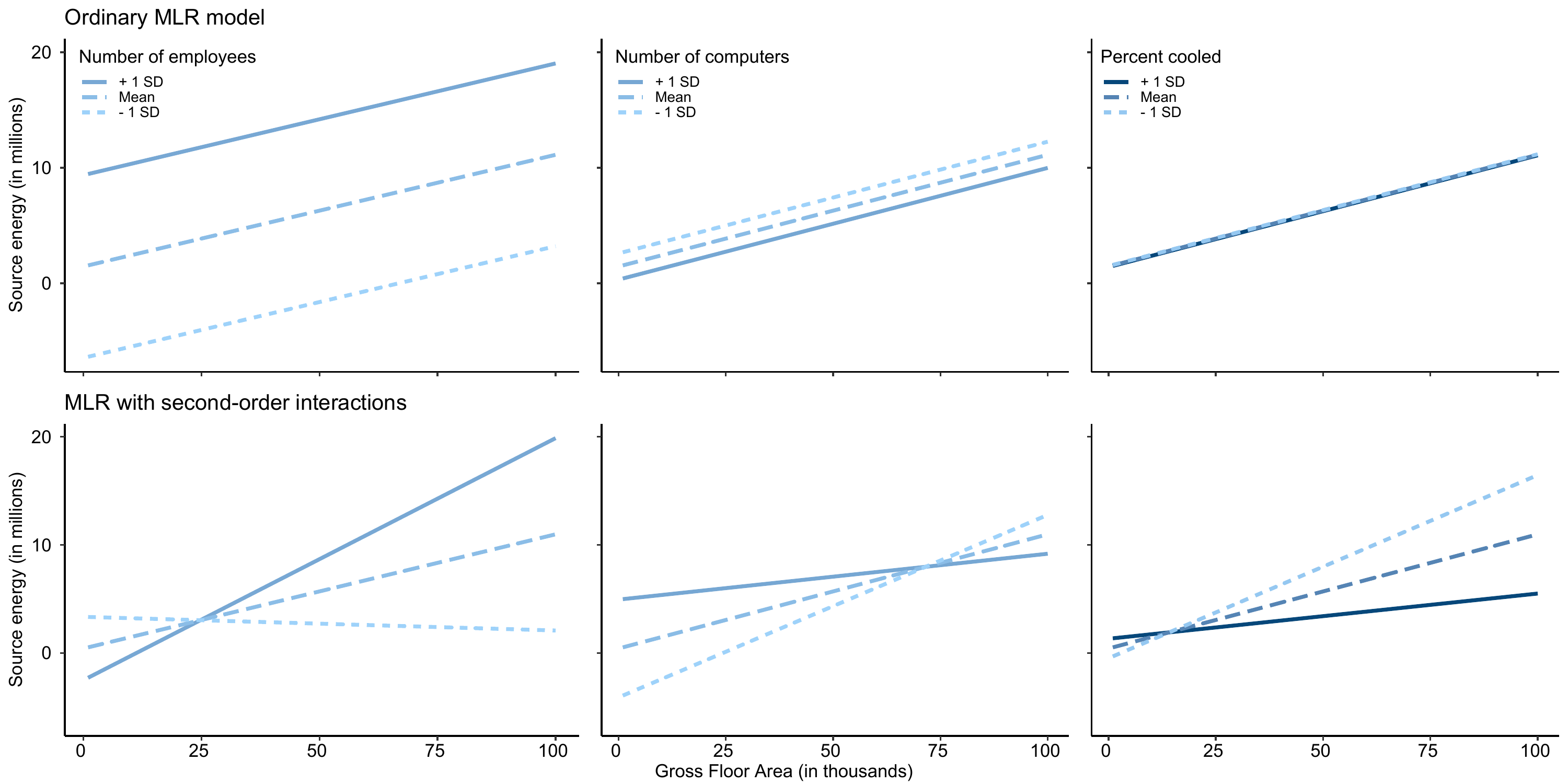}
	\caption{The interaction plots compare the effect of GFA on source energy with respect to three factors, number of employees, number of computers, and CGFA, between ordinary MLR model (top row) and second-order interaction model (bottom row). Office building's energy usage increased with GFA in the ordinary MLR model. In contrast, the interaction model revealed that energy usages actually decreases as GFA increases in some offices where a few number of employees are working.}
	\label{fig:int_office}
	\vspace{5pt}
\end{figure*}

The adjusted $R^2$ value for GBTi2 was more significant than the corresponding values for the other two models for all building types, except worship facilities and K-12 schools. GBTi2 failed on worship buildings because it contained only 69 samples in the Seattle data set, which is far less than 249 samples in the CBECS data set. As an ensemble method, GBTi2 requires more training samples than MLR. Between ordinary and interaction MLR models, MLRi2 performs better than MLR.

The predictive errors were compared using MAPE, as shown in Figure~\ref{fig:all_mape}. The MAPE values were lower by 50\% for hotels, K-12 schools, and multifamily housings when using the CBECS data set. However, MAPE values in offices were very high, at 129\%. However, when interaction terms were included in the model, the MAPE values reduced to 70\%. The high error rate and low adjusted $R^2$ for offices were due to diverse activities within this building type (government offices and courthouses). Adding additional variables specific to each activity or benchmarking them separately could reduce the error rate. Overall, MAPE values were lower with the building-specific data sets than with the CBECS data set for all building types.



\section{Model explainability}
\label{sec:explainability}

The next step in this framework was to explore model explainability. The increased complexity of the proposed models provided a means of capturing behavior that can increase accuracy, but it also provided a method for computing which factor or set of factor interactions was most influencing the prediction. The first focus was on the CBECS data and the implementation of the MLR models. This subsection outlines the process of using the feature interaction and SHAP value visualizations to the user for interpretation capacity. 

\subsection{Interpretation of MLR models with feature interactions}
\label{sec:results:mlr_interpretation}

The groups of variables that were interacting with each other were first analyzed in addition to the significance of these interactions in the model for office buildings. Table~\ref{tab:ols_olsi2} shows the MLR model with second-order interactions when modeling source energy usage. It can be seen that there were nine interaction terms that were statistically significant ($p < 0.01$). This revealed that there were significant interactions among building attributes whose inclusion helps improve model performance. Figure~\ref{fig:int_office} shows the interactions between the gross floor area and the three most influential building attributes: number of employees, number of computers, and operational hours per week. It was observed that the total energy use of office buildings increased linearly with GFA in the ordinary MLR model. When their combined effects were studied using the feature interactions, the model revealed that the effect of GFA on energy usage was influenced by other building attributes. For example, energy usage decreased linearly with GFA in offices where less number of employees were working.



\begin{table}[ht] \centering 
    \ttfamily
    \small
    \begin{tabular}{@{\extracolsep{5pt}}lD{.}{.}{-3} } 
        \toprule
        Variables & \multicolumn{1}{c}{Coefficients} \\ 
        \midrule
        GFA & 202.862^{***} \\ 
        OpenHours & -264.161 \\ 
        WorkersCnt & -21,438.320^{****} \\ 
        ComputersCnt & 11,270.410^{***} \\ 
        Bank & -15,712.320 \\ 
        CGFA & -1,825.969 \\ 
        CDD65 & -3.631 \\ 
        GFA $\cdot$ OpenHours & 0.347^{+} \\ 
        GFA $\cdot$ WorkersCnt & 0.172^{****} \\ 
        GFA $\cdot$ ComputersCnt & -0.130^{****} \\ 
        GFA $\cdot$ CGFA & -1.051^{+} \\ 
        OpenHours $\cdot$ WorkersCnt & -48.270^{***} \\ 
        OpenHours $\cdot$ ComputersCnt & 23.198^{**} \\ 
        WorkersCnt $\cdot$ Bank & -4,857.172^{+} \\ 
        WorkersCnt $\cdot$ CGFA & 268.347^{****} \\ 
        WorkersCnt $\cdot$ CDD65 & -1.083^{*} \\ 
        ComputersCnt $\cdot$ Bank & 2,994.012^{*} \\ 
        ComputersCnt $\cdot$ CGFA & -120.023^{***} \\ 
        ComputersCnt $\cdot$ CDD65 & 0.767^{*} \\ 
        Constant & 40,679.820 \\ 
        \midrule
        Adjusted R$^{2}$ & 0.718 \\ 
        F Statistic & 79.711\\ 
        \hline 
        \hline \\[-1.8ex] 
        \multicolumn{2}{r}{\scriptsize \textit{Note:} + p<0.1; * p<0.05; ** p<0.01; *** p<0.001; **** p<0.0001} \\
    \end{tabular} 
    \caption{MLR model with second-order interaction terms.} 
	\label{tab:ols_olsi2} 
\end{table} 


 \subsection{Interpretation of XGBoost models using SHAP values}
\label{sec:results:xgb_interpretation}

Decision tree-based models inherently capture higher-order interactions among variables. All the variables that appear together on the traversal path from the root to the leaf node interact with each other. The height of a decision tree represents the maximum order of interactions used in the tree. A GBT model with many decision trees will have several interaction terms that are formed by aggregating all feature interactions in each tree. For example, the GBT shown in Figure~\ref{fig:gbt} has four decision trees, each with height two. The building attribute referred to as the percentage total cooled area (\emph{CooledPercent}) was interacting with gross floor area (\emph{m\textsuperscript{2}}) and total operating hours per week (\emph{OpenHours}) in Tree 2. There were two interactions in each tree and a total of eight interactions that can be seen in this sample GBT model.

\begin{figure}
	\centering
	\small
	\includegraphics[scale=0.5]{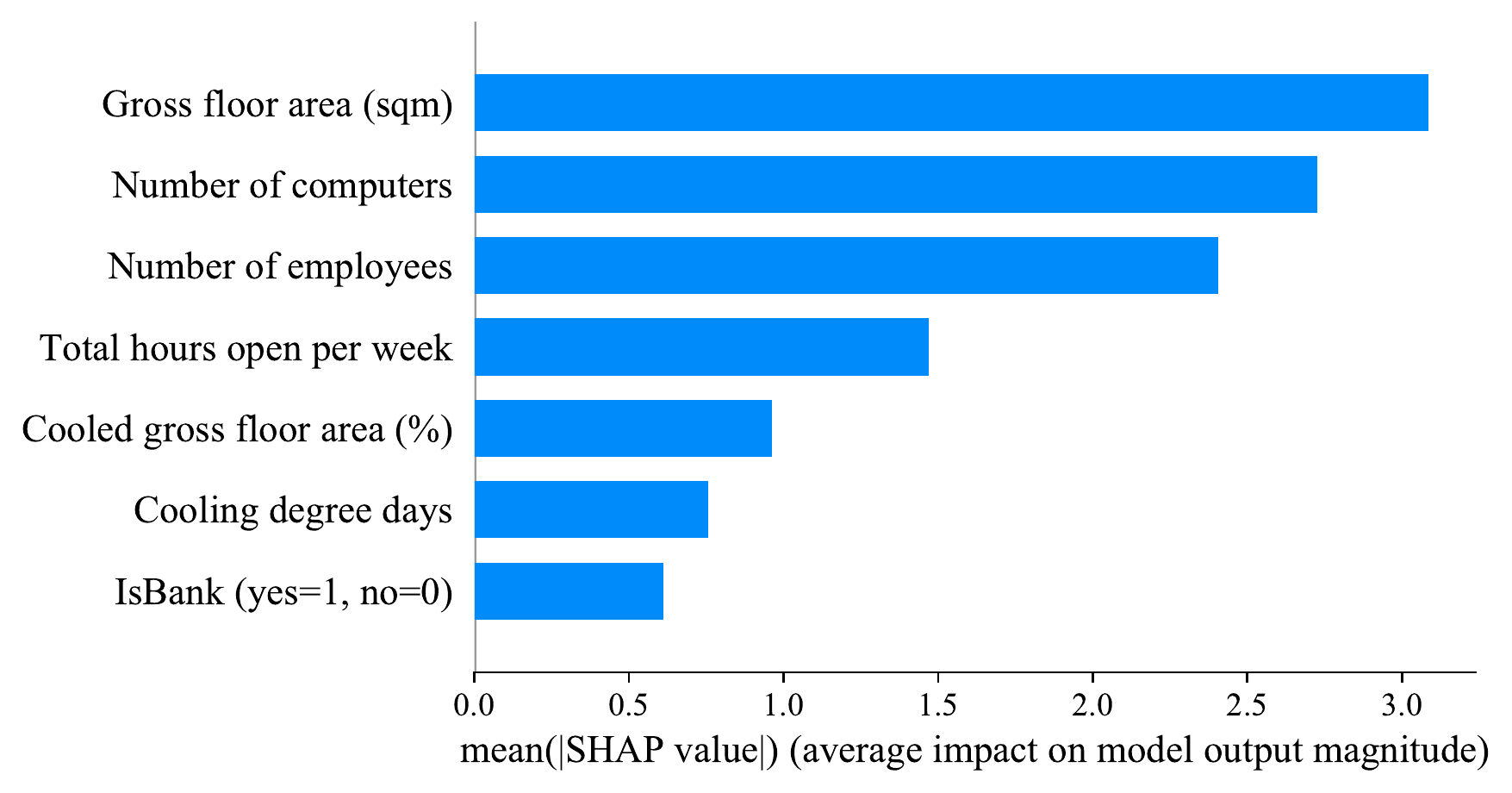}
	\caption{The traditional feature importance plot for office buildings. Feature importance values were calculated by averaging the SHAP values of each attribute.}
	\label{fig:cbecs_gbt_shap_fi}
\end{figure}

\begin{figure}
	\centering
	\small
	\includegraphics[scale=0.5]{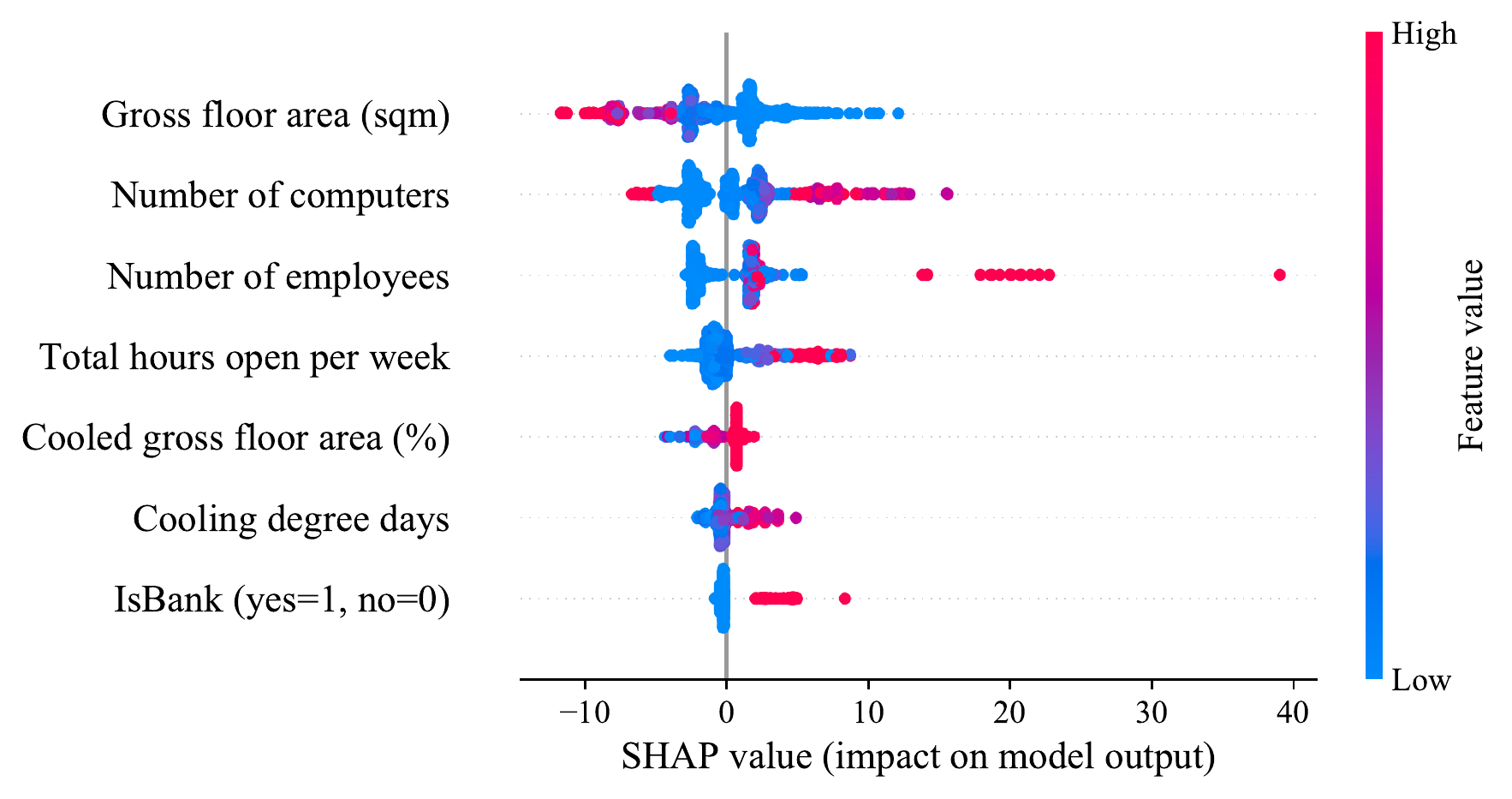}
	\caption{The SHAP summary plot shows the feature importance and their influence on energy usage for office buildings.}
	\label{fig:cbecs_gbt_shap}
\end{figure}

As explained in Section~\ref{sec:shap}, SHAP values were used for interpreting GBT models. The feature importance of a GBT model for office building is shown as a traditional bar chart in Figure~\ref{fig:cbecs_gbt_shap_fi} and as a \emph{SHAP summary plot} in Figure~\ref{fig:cbecs_gbt_shap}. Unlike a traditional bar chart, this summary plot shows the SHAP values of every feature and for every sample. In effect, this is a set of scatter plots, one for each feature, stacked by their order of importance. The y-axis refers to variable names in decreasing order of importance, and the x-axis indicates the SHAP values for each feature ordered from lowest to highest. Each dot represents a sample in the data set, and its gradient color indicates the original value for that feature. Unlike the linear MLR models in which the coefficients denoted the average influence on energy usage, the SHAP values revealed each attribute's influence on energy usage per building level. Similar to the traditional feature importance plot, as shown in Figure~\ref{fig:cbecs_gbt_shap_fi}, it can be inferred that GFA was the most significant feature followed by the number of computers, number of occupants, and average weekly operating hours.

\begin{figure}[ht!]
	\centering
	\small
	\includegraphics[scale=0.7]{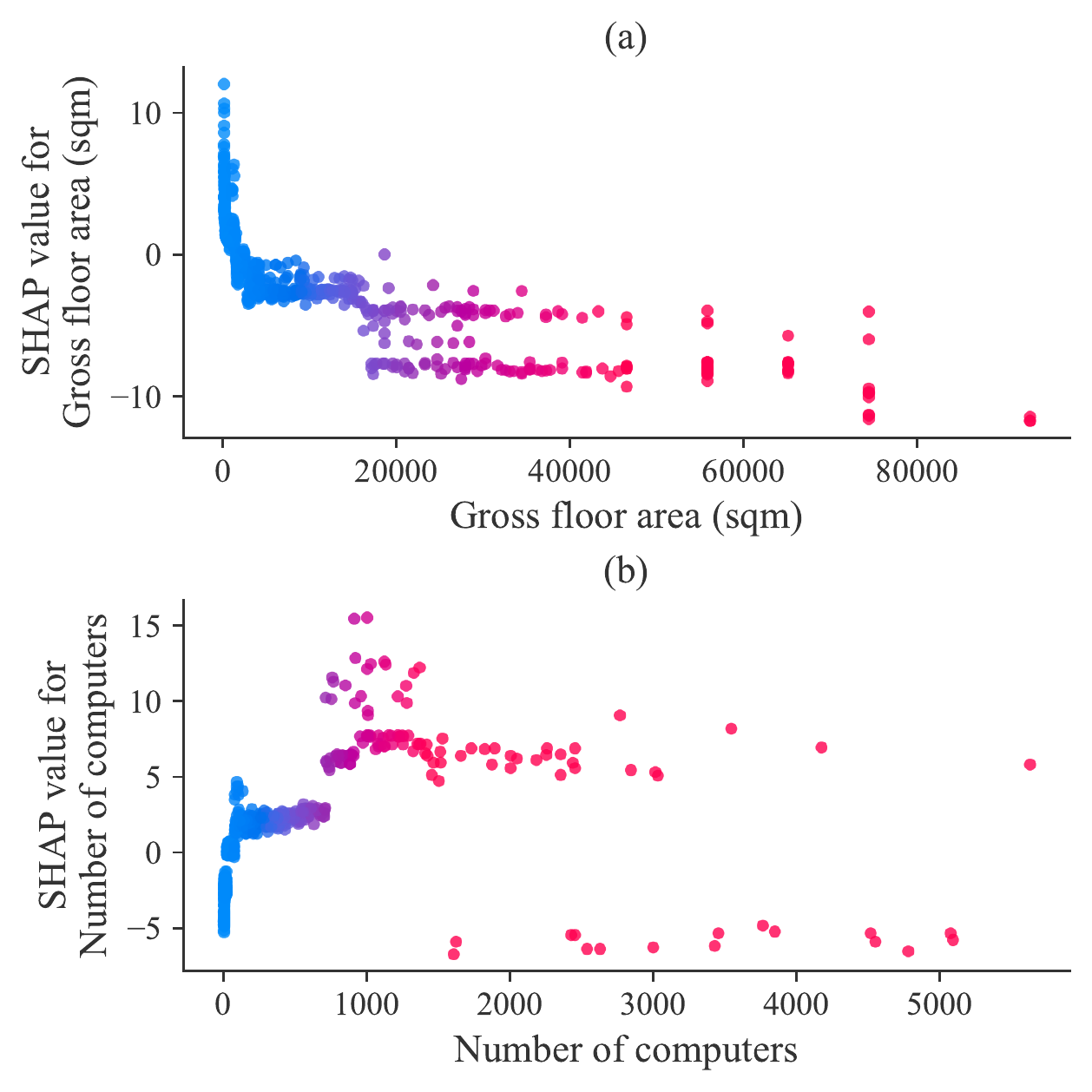}
	\caption{The SHAP dependence plot for GFA and number of computers. Both GFA and number of computers had a relatively linear influence on energy use in small buildings (GFA $<$ 2,415 m\textsuperscript{2} (26,000 ft\textsuperscript{2})) and nonlinear influence in large buildings.}
	\label{fig:cbecs_gbt_shap_dep_sqft}
\end{figure}

\begin{figure}[ht!]
	\centering
	\includegraphics[scale=0.6]{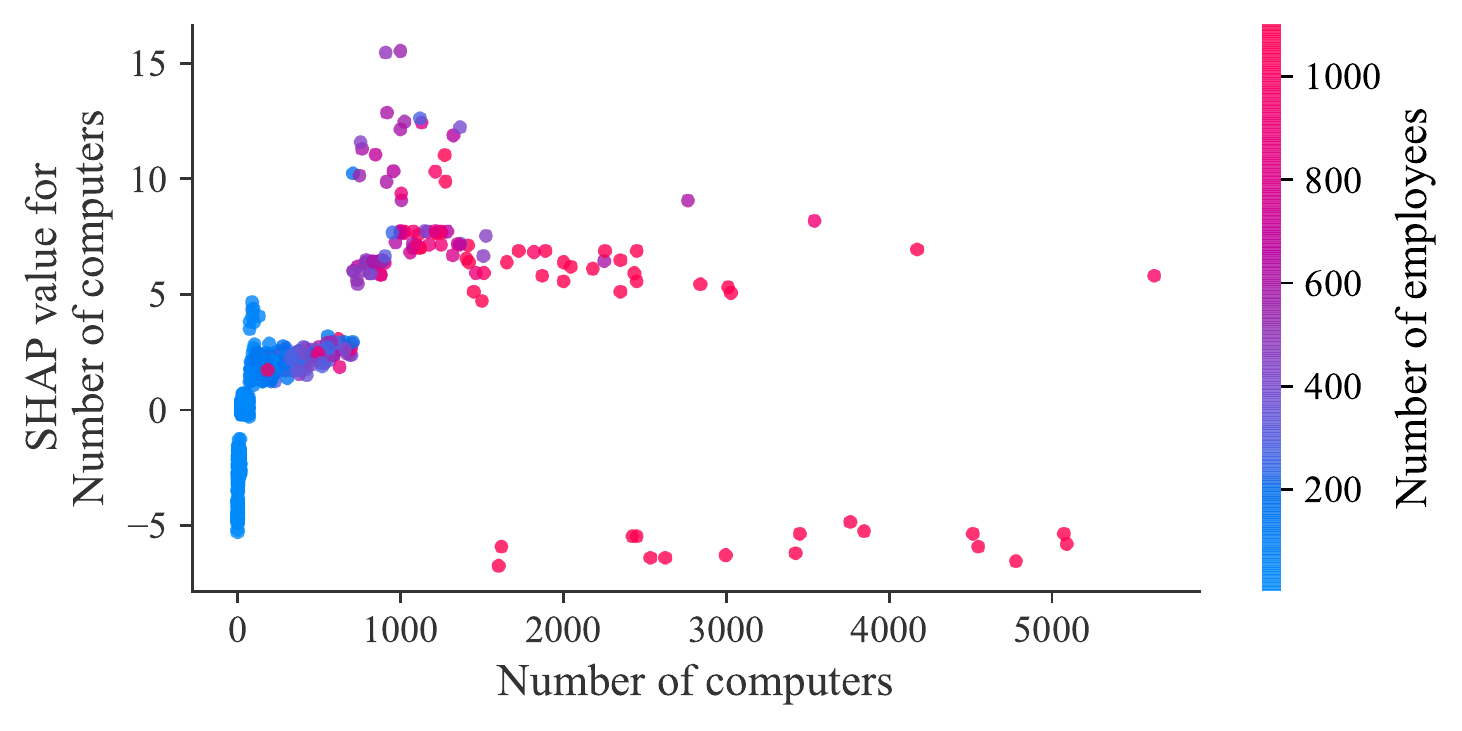}
	\caption{The SHAP dependence plot shows that effect of number computers on model output with respect to number of employees.}
	\label{fig:cbecs_gbt_shap_dep}
\end{figure}

SHAP dependence plots were used to interpret the effect of each building attribute on energy usage. As an example, Figure~\ref{fig:cbecs_gbt_shap_dep_sqft}a shows the dependence plot of a scatter plot between the GFA (x-axis) and the SHAP values of the GFA (y-axis). Each dot represents a building, and its gradient color corresponds to the original value of GFA from low (blue) to high (red). Since SHAP values represent a feature's responsibility for a change in the model output, this dependence plot shows the change in predicted energy use as GFA changes. It was observed that the SHAP values for GFA were positive in small offices (GFA $<$ 2,415 m\textsuperscript{2} (26,000 ft\textsuperscript{2})), as seen from the set of blue dots in the top-left region of the plot. This behavior can be interpreted as GFA had a positive influence on predicted energy in small offices, whereas it had a negative influence on larger buildings (GFA $>=$ 2,415 m\textsuperscript{2} (26,000 ft\textsuperscript{2})). The broader spread of negative SHAP values for large buildings indicates the presence of feature interactions in the model.

Similarly, Figure~\ref{fig:cbecs_gbt_shap_dep_sqft}b shows the dependence plot for number of computers. It was observed that the SHAP values for a number of computers were negative in offices that use few computers ($<$ 67) as seen from the set of blue dots in the lower-left region of the plot. This scenario can be interpreted as the number of computers had a negative influence on predicted energy, whereas it had a positive influence on other buildings that use a large number of computers ($>=$ 67). The wider spread of positive SHAP values indicates the presence of feature interactions in the model.

Figure~\ref{fig:cbecs_gbt_shap_dep_sqft}b also revealed a subset of buildings, as seen from the red dots in the bottom of the plot, in which the number of computers had a negative influence (SHAP values $<$ -60)) on predicted energy despite using a large number of computers (1,603 $<=$ number of computers $<=$ 5,870). This situation is a clear indication of feature interaction effect because other feature(s) dominated the prediction energy in this particular group of buildings. Further investigation revealed that the number of workers was the dominant feature within this subset of buildings. This insight was observed from Figure~\ref{fig:cbecs_gbt_shap_dep} that shows the effect of the number of computers on predicted energy usage with respect to the number of workers. In this plot, the gradient color of each dot corresponds to the original value of the number of workers from low (blue) to high (red). This visualization can be interpreted as the number of workers had a negative influence on predicted energy within this subset of buildings as all buildings within this subset had a more substantial number of employees ($>$ 1,900).

 \begin{figure*}[ht!]
	\centering
	\includegraphics[scale=0.6]{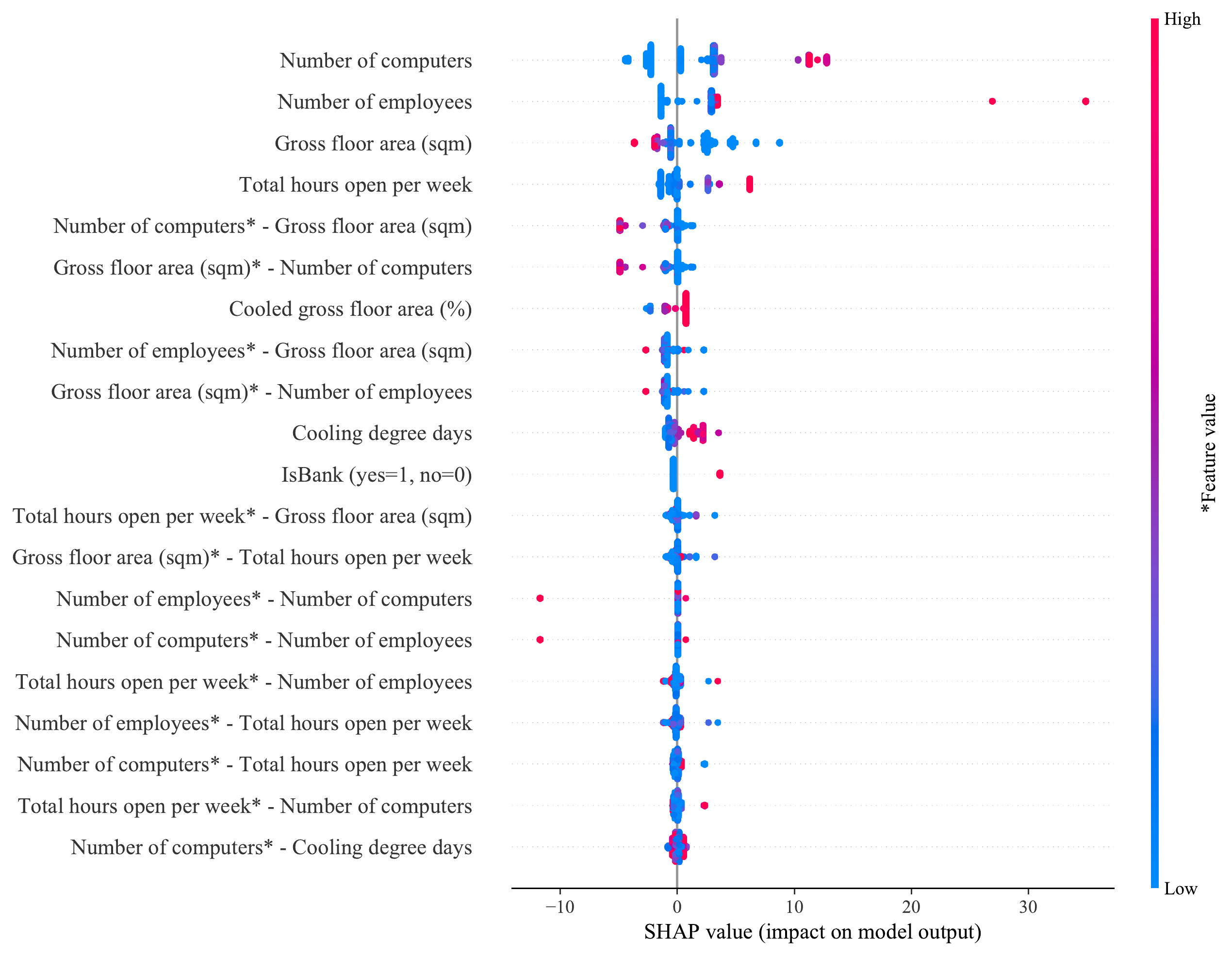}
	\caption{SHAP summary plot shows the feature importance of second order interaction model for office buildings.}
	\label{fig:cbecs_gbt_shap_int}
\end{figure*}

The SHAP values of all second-order interactions, along with main effects, are shown in Figure~\ref{fig:cbecs_gbt_shap_int}. It is seen in this plot, the top three interactions were \emph{GFA:ComputersCnt}, \emph{GFA:WorkersCnt}, and \emph{GFA:OpernHours}. One will note that the MLRi2 model also revealed that these three interaction terms were significant, as shown in Table~\ref{tab:ols_olsi2}. One can perform a similar analysis and identify how does the predicted energy change with respect to change each feature using the SHAP dependence plots.


\begin{figure*}[h]
    \centering
    \small

    \subfloat[This office had a lower predicted EUI (11.8) than its peer group average (16.9), and therefore, it is an \emph{energy-efficient building}. This force plot also reveals which characteristics of the building make it energy efficient. This building has a GFA of 919.7 m\textsuperscript{2} (9,900 ft\textsuperscript{2}) and it is fully air-conditioned (CGFA = 100\%). These two factors push this building to consume more energy. Despite this force (red bars pushing towards the right) there is a much larger set of factors pushing this building to use lower energy (blue bars pushing towards left) because it has a lower number of workers (17), lower number of computers (17), and operates for a lower number of hours (40 hours per week) than its peer group average (See Table~\ref{tab:shap_values}).]{%
       \includegraphics[scale=0.5, trim=0.1cm 3cm 0.1cm 0.1cm]{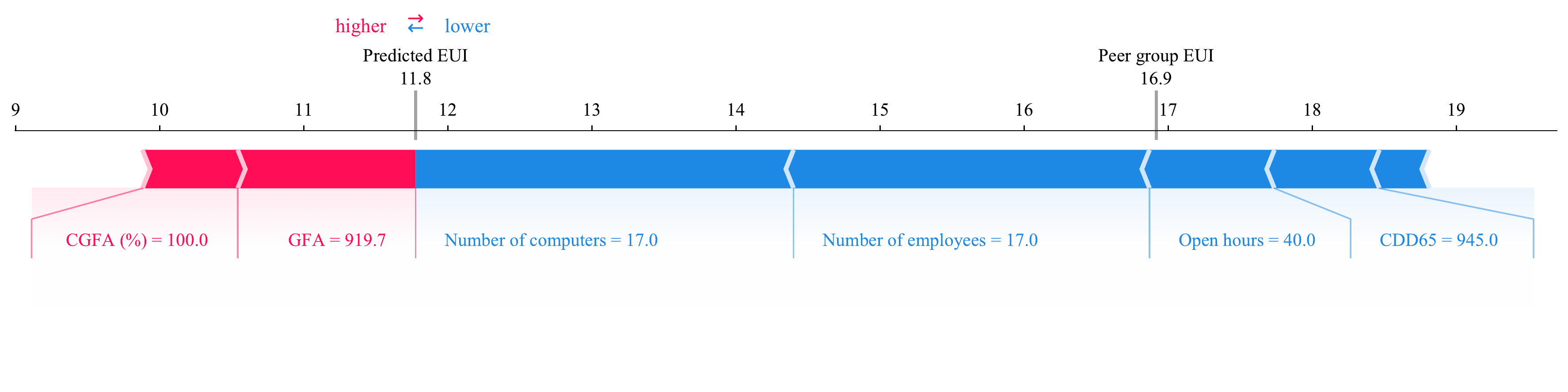}
    }
    
    \subfloat[
    The predicted EUI of this office (24.5) was higher than the peer group average (16.9), and therefore, is an \emph{energy-inefficient building}. The force plot can be used to identify which characteristics of the building made it energy inefficient. This building has a lower number of employees (40) than the peer group average of 234, which pushes this building to consume less energy. Despite this force, there was a much larger set of factors pushing this building to consume more energy because this building operated for longer hours (168), had fully air-conditioned floor space, and was using a lower number of computers (35) than its peer group average (See Table~\ref{tab:shap_values}). Note that using the lower number of computers in a lower GFA (1,347.1 m\textsuperscript{2} (14,500 ft\textsuperscript{2})) office caused the building to use more energy.]{%
        \includegraphics[scale=0.5, trim=0.1cm 3cm 0.1cm 0.1cm]{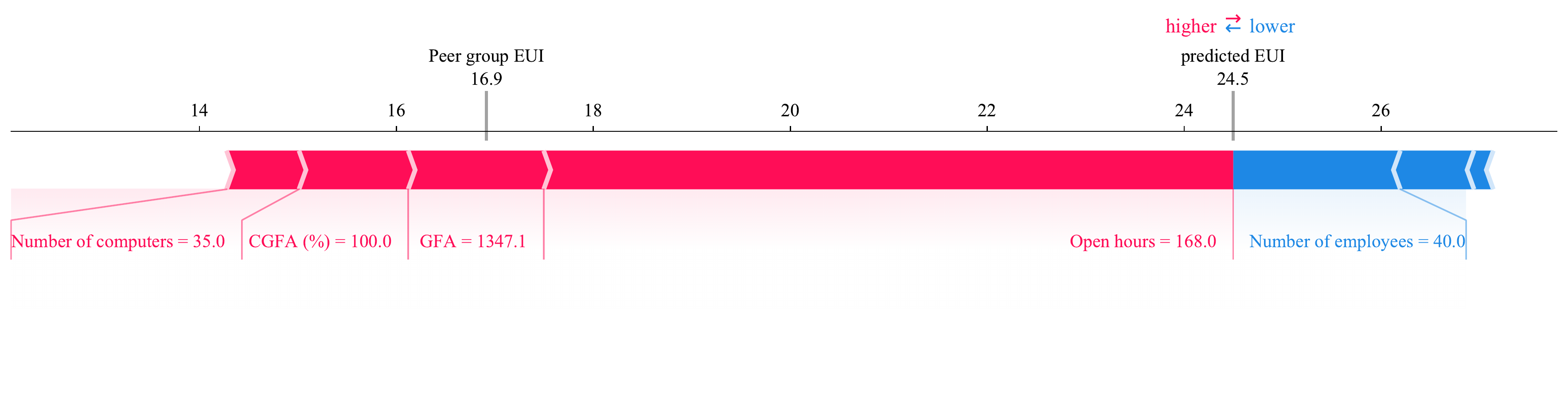}
    }

    \caption{Example SHAP force plots for two office buildings with lower (a) and higher (b) predicted EUI using the GBT model. The width of each bar denotes the SHAP value for that factor, and the color indicates the ones that were influencing the prediction to go higher (red) or lower (blue). Each feature and its value are shown below each SHAP value bar.}
    \label{fig:force}
\end{figure*}


\subsection{SHAP force plot}
\label{sec:results:xgboost_shap}
While the interaction and SHAP visualizations are informative, they are too technical to be understood by the average building performance analyst. The \emph{SHAP force chart} was developed to address this shortcoming \cite{lundberg_explainable_2018}. It illustrates how the input features impact predicted output values. For example, Figure~\ref{fig:force} shows a force plot for a single building in the CBECs data set. This chart illustrates a range of energy consumption, and there are two annotations: one for the \emph{base value}, which is the average model output over the training data set, and the second for the \emph{output value}, which is the energy prediction from the model. This visualization revealed the features that were responsible for the discrepancy between base value and model output. Features pushing the prediction higher are shown in red, and those pushing the prediction lower are in blue. This chart has the potential to inform decision-makers about the key attributes responsible for a building's rating. 

\begin{table*}[ht]
    \centering
    \footnotesize
    \begin{tabular}{l>{\centering}*{7}{r}}
        \toprule
        Building Id & GFA (m\textsuperscript{2}) & OpenHours & WorkersCnt & ComputersCnt & IsBank & CGFA (\%) & CDD65 \\ 
        \midrule
        & \multicolumn{7}{c}{Feature values}\\
        \cmidrule{2-8}
        B1 (low EUI)  & 919.7 & 40 & 17 & 17  & No  & 100  & 945  \\
        B2 (high EUI) & 1347.1 & 168 & 40 & 35  & No  & 100  & 928  \\ 
        Peer group mean & 9658.4 & 62.3 & 234.3 & 330.0 & 8.5\% & 90.4 & 1650.1 \\
        \cmidrule{2-8}
        & \multicolumn{7}{c}{SHAP values}\\
        \cmidrule{2-8}
        B1 (low EUI)  & 1.23 & -0.87 & -2.47 & -2.62 & -0.35 & 0.66 & -0.73 \\
        B2 (high EUI) & 1.38 & 7.00   & -1.70 & 0.74 & -0.20 & 1.11 & -0.75 \\
        \bottomrule
    \end{tabular}
    \caption{Building attributes and their corresponding SHAP values for two office buildings.} 
    \label{tab:shap_values}
\end{table*}


\section{Discussion}
\label{sec:discussion}

This paper has presented two modeling techniques that could increase the effectiveness of benchmarking and rating systems as well as a means of interpreting interactions that influence the end score. This section discusses the potential impact these upgrades to a system like Energy Star could have in the context of various aspects of the built environment.

\subsection{Bridging the gap with asset rating methodologies}
\label{sec:asset_rating}
In many jurisdictions, a component of performance rating systems is what is known as an \emph{asset score} \cite{lee_use_2013}. This score is calculated using the physical and operational parameters of the building to predict how much energy a building should consume in theory. These models are simplified first principles models as opposed to the data-driven models trained from historical measured data outlined in this paper. This type of rating system is exceedingly useful because if the asset rating diverges from the operational rating, then there is a potential for the diagnosis of \emph{why} the building is not performing well. Categories of \emph{diagnosis} include poor operational choices, decreasing system or equipment efficiencies, or bad human behavior. The key downside of the asset rating system is the challenges of data collection of the building's physical attributes, thus the scalability of this method across the building stock. Explainable data-driven modeling can help bridge the gap between these two related types of rating systems by providing another layer of information that can be used to decide whether the asset scores should be calculated in the first place. More accurate and explainable models would provide a \emph{filtering} opportunity for users to decide better whether the next level of analysis is warranted based on the situation. Bridging this analysis gap could improve the scalability and cost-effectiveness of the whole process. 

\subsection{Impact on decision-making behavior related to energy efficiency upgrades or investments}
\label{sec:retrofits}

A world-wide review of building rating schemes touts numerous benefits for such systems around the world. However, one of the drawbacks that were observed and studied was the impact on human behavior and decision-making. While benchmarking studies were able to influence decisions related to ownership transfer, they had minimal impact on energy savings implementation \cite{international_partnership_building_2014}. The objective of influencing the energy savings intervention techniques could extend the ability of rating systems to save energy even more. Explainability of the data-driven models that actual energy consumption is being compared to could provide more information to the decision-maker of the building in terms of energy savings opportunities. For example, if a building owner understands that the model output prediction value their building is being compared to is mostly influenced by the model input for a number of people, then they could plan to install occupancy detection systems in their building to assist in the quantification of energy wastage. An owner might find that their assumed occupancy value is much different from reality, which would be a feedback loop to operations decisions. A dashboard could be developed using the explainability techniques outlined in this paper to showcase a \emph{details-on-demand} style of data visualization. These types of visual analytics innovations would also push the whole community forward in terms of understanding how and why buildings perform well or not.

\subsection{Increased trust in performance rating due to explainable quality assessment}
\label{sec:quality}

As discussed in Section \ref{sec:introduction}, the medical field has tested the impact of explainability on the level of trust that a doctor has in a diagnosis suggestion from a machine learning model. This increase in trust can provide higher adoption of such suggestions \cite{caruana_intelligible_2015}. For buildings, the decision-makers might be more likely to follow the guidance provided by a rating or benchmarking system if they understand how the model has calculated the end values. This visibility can also add a dimension of quality control to the process, where engineering intuition can help understand whether there are significant errors in model input or execution. Once again, visual analytics innovations could further facilitate progress in this area as customized dashboards could provide a more detailed perspective of the \emph{clues} as to why a building is not performing well. This type of visualization could inspire confidence in a building energy expert as the model is not a \emph{black box to them anymore}.

\section{Conclusion}
\label{sec:conclusion}

This paper outlines the use of two proposed data-driven modeling techniques that showed an error rate reduction as compared to the status quo of benchmarking with the Energy Star system. Referring back to the initial research questions, it was found that 1) the MRLi and GBT models tested did have lower error rates (7.0\% and 13.7\% respectively) in the prediction of energy consumption, which resulted in a more accurate representation of the discrepancy between actual and predicted consumption, 2) the feature interaction analysis and SHAP value and associated visualizations provided a window into the inner workings of the prediction models used in this process, and 3) the methodology was accurate and implementable on any large building energy data set from different locations.

It was mentioned previously in this paper that building energy benchmarking rating systems have had widespread success in terms of deployment and energy savings. It was shown that such systems around the world are attaining energy savings of 3-8\% after at least two years of implementing a program. While these savings are a good start, the innovations in this study add the potential for even more energy savings based on the opportunity achieve some of the insight than a physics-based model would provide with far less data collection, improve the decision-making ability for retrofits and operations decisions, and improve the trust in the values calculated by the rating systems themselves.

\subsection{Limitations and reproducibility}
As with any data-driven study, the generalizability of the techniques and results is only applicable to the input data tested. In this case, the process was applied to three large sets of data from various parts of the United States. It can be speculated that the method is implementable in other countries, but this aspect has not been tested. The framework is reproducible and testable on any of the dozens of other city-wide open disclosure program data sets in the world. The analysis procedures in this paper are available online as R Markdown documents. All the code and collected data sets are released as reproducible open source\footnote{\url{https://github.com/buds-lab/energystar-plus-plus}}. The detailed documentation for reproducing the results are given in the code repository. 

\subsection{Future work}
As mentioned, the deployment of this framework on numerous other city, state, or country-wide contexts would be an important next step in understanding the value of the proposed framework. An exhaustive side-by-side comparison of the models presented here with all of the techniques found in the literature would provide a comprehensive source of insight into what the best methods of benchmarking are for the larger building stock. Another key future effort could focus on the actual visual analytics considerations of explainable machine learning for benchmarking on actual users. A metric of trust enhancement using these methods could be quantified through user testing studies of operations professionals and decision-makers who use these rating systems.


\section{Acknowledgement}
This work was supported by the Republic of Singapore’s National Research Foundation (NRF) through a grant to the Berkeley Education Alliance for Research in Singapore (BEARS) for the Singapore–Berkeley Building Efficiency and Sustainability in the Tropics (SinBerBEST) Program. 

\section{Author Credit Statement}
Pandarasamy Anjunan: Conceptualization, Methodology, Data curation, Formal analysis, Software, Writing - original draft. Kameshwar Poolla: Conceptualization, Writing – Review \& Editing. Clayton Miller: Conceptualization, Investigation, Methodology, Data curation, Writing - original draft, Supervision.

\bibliographystyle{model1-num-names}
\bibliography{00_main}

\clearpage

\onecolumn

\appendix 

\section{Model hyper-parameters}

\begin{table*}[ht!]
	\centering
\begin{tabular}{llllr}
\toprule
Model name    & Method$^{*}$ & Parameter        & Description                    & Search range \\
\midrule
\multirow{5}{*}{XGBoost}                    & \multirow{5}{*}{xgbtree}   & max depth & Max Tree Depth            & 2-3 \\
              &        & nrounds          & \# of Boosting Iterations         & 1-200        \\
              &        & eta              & Shrinkage                      & 0.1-0.9      \\
              &        & colsample bytree & Subsample Ratio of Columns     & 0.2-0.8      \\
              &        & subsample        & Subsample Percentage           & 0.25-1       \\
\midrule 
\multirow{3}{*}{SVM-Poly}                    & \multirow{3}{*}{svmPoly}   & degree & Polynomial degree            & 1-3 \\
&        & scale          & Scale         & 0.001 - 1.0       \\
&        & c              & Cost          & 0.25 - 1.0      \\
\midrule
\multirow{2}{*}{SVM-RBF}                    & \multirow{2}{*}{svmRadial}   & sigma & Sigma            & 0.01-1 \\
&        & c              & Cost          & 0.25 - 1.0      \\
\midrule                           
Random Forest & rf     & mtry             & \# of Randomly Selected Predictors & 2-($p$/3)  \\
\midrule
CART          & rpart2 & maxdepth         & Max Tree Depth                 & 2-3            \\
\midrule
\multirow{3}{*}{Neural Network} & \multirow{3}{*}{neuralnet} & layer1    & \# of Hidden Units in Layer 1 &  1-5   \\
              &        & layer2           & \# of Hidden Units in Layer 2      &   1-5           \\
              &        & layer3           & \# of Hidden Units in Layer 3      &    1-5         \\
\bottomrule

\end{tabular}
    \caption{List of models and their hyper-parameters. All models were validated using a 10-fold cross validation method with two repeated rounds to select the final model parameters for each building type. $^{*}$Method denotes the model identifier value passed to the \emph{train} function of \emph{caret} package in R. $p$ denotes the total number of predictors.} 
    \label{tab:model-param}
\end{table*}

\section{Energy Star variables}
\begin{table*}[ht!]
    \centering
    \footnotesize
    \begin{tabular}{l>{\centering}*{6}{c}}
        \toprule
        List of variables & Hotel & K-12 School & Multifamily & Office & Retail & Worship \\ 
        \midrule
        Number of guest rooms per 92.9 m\textsuperscript{2} (1000 ft\textsuperscript{2}) & \checkmark &  &  &  &  &  \\ 
        Number of workers per 92.9 m\textsuperscript{2} (1000 ft\textsuperscript{2}) & \checkmark & \checkmark &  & \checkmark & \checkmark &  \\ 
        Number of refrigeration/freezer units per 92.9 m\textsuperscript{2} (1000 ft\textsuperscript{2}) & \checkmark &  &  &  & \checkmark &  \\ 
        Heating Degree Days x percent of the building that is heated & \checkmark & \checkmark &  &  & \checkmark$^{*}$ & \checkmark \\ 
        Cooling Degree Days x percent of the building that is cooled & \checkmark & \checkmark &  & \checkmark$^{*}$ & \checkmark$^{*}$ & \checkmark \\ 
        Presence of a commercial/large kitchen (yes/no) & \checkmark &  &  &  &  &  \\ 
        Whether there is energy used for cooking (yes/no) &  & \checkmark &  &  &  &  \\ 
        Whether the school is open on weekends (yes/no) &  & \checkmark &  &  &  &  \\ 
        Whether the school is a high school (yes/no) &  & \checkmark &  &  &  &  \\ 
        Number of units per 92.9 m\textsuperscript{2} (1000 ft\textsuperscript{2}) &  &  & \checkmark &  &  &  \\ 
        Number of bedrooms per unit &  &  & \checkmark &  &  &  \\ 
        Total Heating Degree Days (base 18.3\ensuremath{^\circ}C (65\ensuremath{^\circ}F)) &  &  & \checkmark &  &  &  \\ 
        Total Cooling Degree Days (base 18.3\ensuremath{^\circ}C (65\ensuremath{^\circ}F)) &  &  & \checkmark &  &  &  \\ 
        Low-rise building (yes/no) &  &  & \checkmark &  &  &  \\ 
        Gross floor area (m\textsuperscript{2}) &  &  &  & \checkmark &  &  \\ 
        Weekly operating hours &  &  &  & \checkmark & \checkmark & \checkmark \\ 
        Number of computers per 92.9 m\textsuperscript{2} (1000 ft\textsuperscript{2}) &  &  &  & \checkmark &  &  \\ 
        Whether or not the building is a bank branch (yes/no) &  &  &  & \checkmark &  &  \\ 
        Whether the building is a supermarket (yes/no) &  &  &  &  & \checkmark &  \\ 
        Adjusted for number of workers per 92.9 m\textsuperscript{2} (1000 ft\textsuperscript{2}) for supermarket &  &  &  &  & \checkmark &  \\ 
        Number of religious worship seats per 92.9 m\textsuperscript{2} (1000 ft\textsuperscript{2}) &  & &  &  &  & \checkmark \\ 
        Percent of gross floor area used for food preparation &  &  &  &  &  & \checkmark \\ 
        \midrule
        Total number of variables & 6 & 6 & 5  & 6 & 7 & 5 \\ 
        \bottomrule
        $^*$Using natural log of Cooling/Heating Degree Days &  &  &  &  &  & \\ 
    \end{tabular}
    \caption{List of building attributes used in the Energy Star system for different building types.} 
    \label{tab:energystar_vars}
\end{table*}


\end{document}